\documentclass[a4paper,11pt]{article}	
\pdfoutput=1 
\usepackage{jheppub} 

\usepackage[table]{xcolor}                  
\usepackage{colortbl}
\usepackage{booktabs}                       
\usepackage{multirow,bigdelim}              
\usepackage{longtable}                      
\usepackage{arydshln}                       

\usepackage{units}                          
\usepackage{amsmath,amssymb,mathtools}      
\usepackage{slashed}                        
\usepackage{amsthm}
\usepackage{bbm,dsfont}                     

\usepackage{bm}                             
\usepackage{upgreek}                        
\usepackage[vcentermath]{youngtab}          
\usepackage{ytableau}                       
\usepackage{empheq}                         
\usepackage{suffix}	                        
\usepackage{lscape}	                        

\usepackage{graphicx}                       
\usepackage{subfig}	                        
\usepackage{tikz}                           
\usepackage{float}
\usepackage[export]{adjustbox}              

\usepackage{url}                            
\usepackage{makeidx}                       
\usepackage[numbers,sort&compress]{natbib}  
\usepackage[colorlinks=true,urlcolor=blue,anchorcolor=blue,citecolor=blue,filecolor=blue,linkcolor=blue,menucolor=blue,pagecolor=blue,linktocpage=true,pdfproducer=medialab,pdfa=true]{hyperref}	

\usepackage{ifthen}	                        

\allowdisplaybreaks[2]          

\graphicspath{{figs/}}

\def\be{\begin{equation}}
\def\ee{\end{equation}}

\def\bea{\begin{eqnarray}}
\def\eea{\end{eqnarray}}

\makeindex

\title{\boldmath Wilson-loop One-point Functions in ABJM Theory }
\author[a]{Yunfeng Jiang}
\author[b,c]{Jun-Bao Wu\footnote{Corresponding author}}
\author[b]{Peihe Yang}
\affiliation[a]{School of Phyiscs and Shing-Tung Yau Center\\
  Southeast University, Nanjing 210096, China}
\affiliation[b]{Center for Joint Quantum Studies and Department of Physics, School of Science,\\ Tianjin University, 135 Yaguan Road, Tianjin 300350, P. R. China}
\affiliation[c]{Peng Huangwu Center for Fundamental Theory, Hefei, Anhui 230026, P. R. China}

\preprint{CJQS-2023-nnn, USTC-ICTS/PCFT-23-11}
\emailAdd{jinagyf2008@gmail.com}
\emailAdd{junbao.wu@tju.edu.cn}
\emailAdd{peihe$\_$yang@tju.edu.cn}

\abstract{In this paper we initiate the study of correlation functions of a single trace operator and a circular  supersymmetric Wilson loop in ABJM theory. The single trace operator is in the scalar sector and is an eigenstate of the planar two-loop dilatation operator. The Wilson loop is in the fundamental representation of the gauge group or a suitable (super-)group.  Such correlation functions at tree level can be written as an overlap of the Bethe state corresponding to the single trace operator and a boundary state which corresponds to the Wilson loop. There are various type of supersymmetric Wilson loops in ABJM theory. We show that some of them correspond to tree-level integrable boundary states while some are not. For the  tree-level integrable ones, we prove their integrability and obtain analytic formula for the overlaps. For the non-integrable ones, we give examples of non-vanishing overlaps for Bethe states which violate selection rules.}

\begin{document}
\maketitle
\flushbottom

\section{Introduction}
\label{sc:intr}

Integrable structure  of four-dimensional $\mathcal{N}=4$ super Yang-Mills (SYM) theory enables us to compute many physical observables non-perturbatively in the planar limit.\footnote{For a collection of reviews, see~\cite{Beisert:2010jr}.} The study of integrability in AdS/CFT was initiated by the discovery that the planar one-loop dilatation operator in the scalar sector is identical to the Hamiltonian of an  integrable  spin chain~\cite{Minahan:2002ve}. Later this result was generalized to the full sector and all-loop order. In the asymptotic regime, the spectrum of local operators can be computed by the all-loop asymptotic Bethe ansatz, which was first proposed in~\cite{Beisert:2005fw} and derived more rigorously in \cite{Beisert:2005tm}. For operators with finite length, the so-called finite size corrections should be taken into account. To solve this challenging problem, different approaches such as the L\"uscher formula \cite{Bajnok:2008bm} and the thermodynamic Bethe ansatz (TBA)~\cite{Ambjorn:2005wa,Bombardelli:2009ns,Gromov:2009tv,Arutyunov:2009ur} have been developed, and finally culminated in the quantum spectral curve (QSC) method~\cite{Gromov:2013pga}.

Single-trace operators in $\mathcal{N}=4$ SYM are mapped to \emph{closed} spin chain states. It turns out that integrable \emph{open} spin chains also play important roles in AdS/CFT. There are at least two ways that open chains could emerge. The first is by changing the theory. Examples include theories with matters in the fundamental representation of the gauge group, such as four-dimensional $\mathcal{N}=2$ $Sp(N)$ theory
~\cite{Chen:2004mu, Chen:2004yf} and $\mathcal{N}=2$ theory obtained by adding flavors to
$\mathcal{N}=4$ SYM~\cite{Erler:2005nr}. The second way is considering specific objects within ${\mathcal N}=4$ SYM
theory which play the role of integrable boundaries. Such objects include domain walls~\cite{DeWolfe:2004zt}, determinant operators which are dual to giant gravitons~\cite{Berenstein:2005vf} and Wilson lines~\cite{Drukker:2006xg, Correa:2018fgz}.

More recently, the study of domain wall one-point functions in defect $\mathcal{N}=4$ SYM \cite{deLeeuw:2015hxa, Buhl-Mortensen:2015gfd}  introduced integrable boundary states into  AdS/CFT integrability\footnote{In~\cite{DeLeeuw:2018cal}, it was proved that the boundary states from the domain wall in the D3-D5 case satisfy the condition for the integrable boundary states in~\cite{Piroli:2017sei}.}. Integrable boundary states are specific states in the Hilbert space which are annihilated by odd conserved charges. Later integrable boundary states also appear in the computation of the correlation function of two determinant operators and a single trace operator~\cite{Jiang:2019xdz, Jiang:2019zig}, 't~Hooft loop one-point functions~\cite{Kristjansen:2023ysz} and Wilson-loop one-point functions~\cite{JKV}.\footnote{A class of fermionic BPS Wilson loops in four-dimensional $\mathcal{N}=2$ quiver theories and $\mathcal{N}=4$ SYM were constructed  last year~\cite{Ouyang:2022vof}. it is interesting to study whether they lead to integrable open chains and/or integrable boundary states.}  Although the integrable boundary states appear quite naturally in the cases of domain walls, 't~Hooft loops and Wilson loops, their emergence in the correlation functions involving two determinant operators are less obvious. The integrable boundary states only show up after some non-trivial computations, such as using large-$N$ effective field theory or performing partial Wick contractions between the giant gravitons.

Three-dimensional $\mathcal{N}=6$ Chern-Simons-matter theory (ABJM theory)~\cite{Aharony:2008ug} is another important example of supersymmetric gauge theories which are integrable in the planar limit.\footnote{The review~\cite{Klose:2010ki} summarised the related achievement till the end of $2010$.} Compared to $\mathcal{N}=4$ SYM theory, almost every aspect of integrability gets more complicated and challenging, due to its smaller symmetry. Similar to $\mathcal{N}=4$ SYM theory, the first hint of integrability of ABJM theory comes from the fact that the planar two-loop dilatation operator in the scalar sector is integrable~\cite{Minahan:2008hf, Bak:2008cp}. All-loop asymptotic Bethe ansatz equations were proposed in~\cite{Gromov:2008qe}. But there is a to-be-determined interpolating  function $h(\lambda)$ appearing in the dispersion relation of the magnons. A conjecture for the  exact expression of $h(\lambda)$ was proposed~\cite{Gromov:2014eha}, based  on the computation of the planar slope function using QSC~\cite{Cavaglia:2014exa} and the result on the vacuum expectation value of $1/6$-BPS bosonic circular Wilson loop~\cite{Drukker:2008zx, Chen:2008bp, Rey:2008bh} computed using supersymmetric localization~\cite{Kapustin:2009kz, Marino:2009jd}.

Integrable open spin chains are relatively less studied from the perspective of 3d super-Chern-Simons theories. Planar two-loop reflection matrices of  open chains from $\mathcal{N}=3$ flavored ABJM theory  was shown to satisfy boundary Yang-Baxter equations  (BYBEs) in~\cite{Bai:2017jpe}, this is a strong evidence for these chains to be integrable at two-loop. Quite recently, it was proved~\cite{Correa:2023lsm} that all-loop reflection matrices of open chains from half-BPS Wilson lines in ABJM theory satisfy BYBEs, under certain assumptions~\cite{Bargheer, Ahn, Drukker:2019bev}. TBA equations for composite operators inserted in cusped Wilson lines were also obtained~\cite{Correa:2023lsm}.
 Solutions of the TBA equations can be used to  confirm the conjectured interpolating function $h(\lambda)$. As for the open chain from the determinant operators in ABJM theory, its two-loop integrability was checked using coordinate Bethe ansatz (CBA)~\cite{Chen:2018sbp} and proved by algebraic Bethe ansatz (ABA)~\cite{Bai:2019soy}. A proposal for the asymptotic all-loop Bethe ansatz equations was given in~\cite{Chen:2019igg}.
The boundary reflection  exchanges A-type magnons and B-type magnons in the flavored ABJM case, while the type of the magnon is preserved during the reflection in the determinant operator and Wilson line cases.

The study of integrable boundary states in ABJM theory started with the computation of three-point functions involving two determinant operators and a single trace operator~\cite{Yang:2021hrl}. Later it was shown that certain domain walls in ABJM theory also lead to integrable boundary states~\cite{Kristjansen:2021abc}\footnote{It was shown in~\cite{Linardopoulos:2021rfq, Linardopoulos:2022wol} that certain D-branes dual to domain walls in both $\mathcal{N}=4$ SYM and ABJM theory indeed provide the integrable boundary conditions for open string attached to them. }. In both cases, the integrable boundary states satisfy the twisted integrability condition~\cite{Gombor:2020kgu} at two-loop level. The aim of this paper is to initiate the study of correlation functions of a single trace operator and a BPS Wilson loop in ABJM theory. As we will see, this is another important set-up where integrable boundary states emerges naturally.

There are various types of  BPS Wilson loops in ABJM theory (see the review~\cite{Drukker:2019bev}). The first BPS Wilson loops which were constructed are the bosonic  $1/6$-BPS ones~\cite{Drukker:2008zx, Chen:2008bp, Rey:2008bh}. The construction is based on the  bosonic $1/2$-BPS ($1/3$-BPS) Wilson loops in general $\mathcal{N}=2$ ($\mathcal{N}=3$) super-Chern-Simons theory~\cite{Gaiotto:2007qi} and is similar to the half-BPS Maldacena-Wilson loop~\cite{Maldacena:1998im, Rey:1998ik} in $\mathcal{N}=4$ SYM.  These Wilson loops correspond to F-strings smearing in a $\mathbf{CP}^1\subset \mathbf{CP}^3$ in the dual $AdS_4\times \mathbf{CP}^3$ background~\cite{Drukker:2008zx, Rey:2008bh}. In another word, the worldsheet theory has Neumann  boundary conditions for the directions along the  $\mathbf{CP}^1$ subspace~\cite{Lewkowycz:2013laa}. The existence of certain half-BPS probe F-string solutions~\cite{Drukker:2008zx, Rey:2008bh} with Dirichlet boundary conditions in all directions of $\mathbf{CP}^3$
indicates the existence of half-BPS Wilson loops which are invariant under a subgroup $SU(3)\times U(1)$ of $SU(4)_R$. Such Wilson loops were constructed by Drukker and Trancanelli~\cite{Drukker:2009hy}, who introduced fermions in the construction of the Wilson loops. Fermionic $1/6$-BPS Wilson loops were constructed in~\cite{Ouyang:2015iza, Ouyang:2015bmy}, based on the construction of fermionic half-BPS Wilson loops in generic $\mathcal{N}=2$ super-Chern-Simons theories. These fermionic $1/6$-BPS Wilson loops in ABJM theory include  the above half-BPS Wilson loops and bosonic $1/6$-BPS ones as special cases.  A subclass  of these fermionic  $1/6$-BPS Wilson lines are shown to dual to F-strings with complicated mixed boundary conditions~\cite{Correa:2019rdk}. In this paper, we will study the Wilson-loop one-point function within a subclass of circular fermonic $1/6$-BPS Wilson loops,\footnote{More precisely speaking,  this subclass was chosen here to insider the Class I of the classification in~\cite{Ouyang:2015iza, Ouyang:2015bmy}. The situation for Class II should be similar. }  and $1/3$-BPS Wilson loops constructed based on the $1/3$-BPS Wilson  lines in~\cite{Drukker:2022txy}.

We will see that the corresponding structure constant can be calculated as the overlap of a boundary state and a Bethe state. The tree-level computation of this one-point function demands the single-trace operator to be an eigenvector of two-loop dilatation operator, which is identical to an integrable Hamiltonian of an alternating SU(4) spin chain~\cite{Minahan:2008hf, Bak:2008cp}. The eigenvectors can be constructed by Bethe ansatz, with additional zero momentum condition. A particularly interesting question for us is that, which supersymmetric Wilson loops correspond to \emph{integrable} boundary states?

We find that the boundary states  corresponding to generic $1/6$-BPS Wilson loops in this subclass  are not integrable. Only two special cases, namely the bosonic $1/6$-BPS and half-BPS Wilson loops give rise to  tree-level integrable boundary states. This result implies that the boundary states corresponding to $1/3$-BPS Wilson loops are also integrable  at tree level. All the aforementioned  tree-level integrable boundary states satisfy twisted integrability condition which leads to selection rule  $\mathbf{u}=-\mathbf{v}, \mathbf{w}=-\mathbf{w}$ where $\mathbf{u}$ and $\mathbf{v}$ are the two sets of momentum carrying Bethe roots and $\mathbf{w}$ are the auxiliary roots. For  these tree-level integrable boundary states, we obtain analytic formulas for the Wilson loop one-point functions (normalized overlaps)  in terms of Bethe roots up to an unimportant phase factor. 

The remaining part of  this paper is organized as follows. In section~\ref{sec:WL}, we review the construction of various supersymmetric Wilson loops in ABJM theory. In section~\ref{sc:onepoint} we compute the Wilson loop one-point functions and find the circular Wilson loops which correspond to integrable boundary states. We prove the  tree-level integrability of such states by algebraic Bethe ansatz and derive the selection rules for the exact overlap. In section~\ref{sec:onesixWL} and~\ref{sec:onehalfWL}, we derive the exact overlap formula for the bosonic 1/6-BPS and 1/2-BPS Wilson loops respectively. We conclude in section~\ref{sec:conclude} and discuss some future directions. Two appendices include our conventions for ABJM theory and numerical solutions of  the Bethe equations.

\section{Various BPS Wilson loops in ABJM theory}
\label{sec:WL}

In this section, we list  Wilson loops that will be studied in this paper. Among these Wilson loops, the $1/3$-BPS circular Wilson loops are new, and they are constructed based on
$1/3$-BPS Wilson lines in~\cite{Drukker:2022txy}. We consider the ABJM theory in three-dimensional Euclidean space $\mathbf{R}^3$ and adopt the notations in \cite{Ouyang:2015bmy}.
The spinor convention, the Lagrangian and the supersymmetry transformation are listed in Appendix~\ref{sc:lagrangian}.

\paragraph{Bosonic $1/6$-BPS  circular WLs.} These loops were first constructed in~\cite{Drukker:2008zx, Chen:2008bp, Rey:2008bh}. 
We consider the loops along $x^\mu=(R\cos\tau, R\sin\tau, 0), \tau\in [0, 2\pi]$.
The construction is the following,
\begin{align}
W_{1/6}^B=&\,\mathrm{Tr} \mathcal{P}\, \exp\left(-i\oint d\tau \mathcal{A}_{1/6}^B(\tau)\right)\,,\,\,&  \hat{W}_{1/6}^B=&\,\mathrm{Tr} \mathcal{P}\,\exp\left(-i\oint d\tau \hat{\mathcal{A}}_{1/6}^B(\tau)\right)\,,\label{bosonic}\\
\mathcal{A}_{1/6}^B=&\,A_\mu \dot{x}^\mu+\frac{2\pi}{k}R_{I}^{\,\,\,J}Y^I Y^\dagger_J|\dot{x}|\,,\,&
\hat{\mathcal{A}}_{1/6}^B=&\,\hat{A}_\mu \dot{x}^\mu+\frac{2\pi}{k}R_{I}^{\,\,\,J} Y^\dagger_JY^I|\dot{x}|\,,
\end{align}
where $\dot{x}^\mu=\frac{dx^\mu}{d\tau}$, and $R_{I}^{\,\,\,J}=\mathrm{diag}(i, i, -i, -i)$.

These two Wilson loops preserve the same supersymmetries,
\bea &&\label{susy}\vartheta_{12}=i R^{-1}  \gamma_3 \theta_{12}\,\, \vartheta_{34}=-i R^{-1} \gamma_3 \theta_{34}\,, \nonumber\\
&&\theta_{13}=\theta_{14}=\theta_{23}=\theta_{24}=0\,, \vartheta_{13}=\vartheta_{14}=\vartheta_{23}=\vartheta_{24}=0\,.\eea

We can combine the above two connections into a big one,
\bea \label{bigbosonic}
L^B_{1/6}=\left(\begin{array}{cc}
\mathcal{A}_{1/6}^B & \\
&  \hat{\mathcal{A}}_{1/6}^B\\
\end{array}\right)\,,\eea
and construct the following Wilson loops
\be W_{1/6}^{B, \mathrm{big}}=\mathrm{Tr} \mathcal{P}\, \exp \left(-i\oint d\tau L^B_{1/6}(\tau)\right)\,. \ee
Obviously the preserved supersymmetries are still the ones in~(\ref{susy}).

\paragraph{Fermionic $1/6$-BPS circular WLs.}
These loops were constructed in~\cite{Ouyang:2015iza, Ouyang:2015bmy}.
We focus on the Class~I loops according to the classification in these papers. Let us consider the loops along the same contour $x^\mu(\tau)=(R\cos\tau, R\sin\tau, 0)$
as the  bosonic $1/6$-BPS Wilson loops,
\bea && W^F_{1/6}=\mathrm{Tr} \mathcal{P} \, \exp \left(-i\oint d\tau L^F_{1/6}(\tau)\right)\,,\,\quad
L^F_{1/6}=\left(\begin{array}{cc}
    \mathcal{A} & \bar{f}_1  \label{fonesixth}\\
   f_2  &  \hat{\mathcal{A}}
\end{array}\right)\,,\label{fermionic16}\\
&& \mathcal{A}=A_\mu\dot{x}^\mu+\frac{2\pi}{k} U_{I}^{\,\,\,J} Y^I Y^\dagger_J |\dot{x}|\,, \, \,\quad
\bar{f}_1=\sqrt{\frac{2\pi}{k}}\bar{\alpha}^I\bar{\zeta} \psi_I |\dot{x}|\,,\\
&& \hat{\mathcal{A}}=\hat{A}_\mu\dot{x}^\mu+\frac{2\pi}{k} U_{I}^{\,\,\,J}Y^\dagger_J Y^I|\dot{x}|\,,\, \quad f_2=\sqrt{\frac{2\pi}{k}}\psi^{\dagger I}\eta \beta_I|\dot{x}|\,,
\eea
with
\bea
\bar{\alpha}^I&=&(\bar{\alpha}^1, \bar{\alpha}^2, 0, 0)\,,\,\quad
\beta_I=(\beta_1, \beta_2, 0, 0)\,,\\
\bar{\zeta}^\alpha&=&(e^{i\tau/2}, e^{-i\tau/2})\,,\,\quad
\eta_\alpha=\left(\begin{array}{c}
     e^{-i\tau/2}  \\
      e^{i\tau/2}
\end{array}\right)\,,\\
U_{I}^{\,\,\,J}&=&\left(\begin{array}{cccc}
    i-2\bar{\alpha}^2\beta_2 & 2\bar{\alpha}^2\beta_1 &0&0  \\
    2\bar{\alpha}^1\beta_2 & i-2\bar{\alpha}^1\beta_1 &0 &0\\
    0&0 & -i & 0  \\
    0& 0& 0   & -i
\end{array}\right)\,.\label{fconstruction}\eea
The preserved supersymmetric is the same as the ones in~(\ref{susy}) for generic $\bar{\alpha}^I, \beta_I$.

It was later noticed that we have the equivalence relation~\cite{Drukker:2019bev, Drukker:2020opf},
\be \label{equivalence} (\bar{\alpha}^I, \beta_J)\sim (\lambda \bar{\alpha}^I, \lambda^{-1}\beta_J)\,, \, \lambda\in \mathbf{C}^\ast=\mathbf{C}-\{0\}\,.\ee

 One can set $\bar{\alpha}^2=\beta_2=0$ to get a subclass of Wilson loops.  Then in this subclass, we have $U_{I}^{\,\,\,J}=\mathrm{diag} (i, i-2\bar{\alpha}^1\beta_1, -i, -i)$, and
\be  \bar{f}_1=\sqrt{\frac{2\pi}{k}}\bar{\alpha}^1\bar{\zeta}\psi_1|\dot{x}|\,,\,\quad
f_2=\sqrt{\frac{2\pi}{k}}\psi^{\dagger 1}\eta\beta_1|\dot{x}|\,.\ee
 Similar subclass in the fermionic $1/6$-BPS Wilson lines was considered in the study of the dual string theory prescription \cite{Correa:2019rdk}.

\paragraph{Half-BPS circular WLs.}
Half-BPS Wilson loops were first constructed in \cite{Drukker:2009hy}. A class of them,  $W_{1/2}$,  appears among the above class of fermionic $1/6$-BPS Wilson loops when the parameters $\bar{\alpha}^I, \beta_I$ satisfy the following constraints,
\be \beta_I=\frac{i\alpha_I}{\bar{\alpha}^J\alpha_J}\,,   \label{constraints}\ee
and at least one of $\bar{\alpha}^1, \bar{\alpha}^2$ is non-zero.  Here $\alpha_I$ is defined by $\alpha_I=(\bar{\alpha}^I)^\ast$.
The preserved supersymmetries are now enhanced to
\be \bar{\alpha}^I\vartheta_{IJ}=iR^{-1}\gamma_3\bar{\alpha}^I\theta_{IJ}\,, \,\quad
\epsilon^{IJKL}\alpha_J\vartheta_{KL}=-iR^{-1}\gamma_3 \epsilon^{IJKL}\alpha_J\theta_{KL}\,.\ee

Using a suitable R-symmetry transformation  acting only on $I=1, 2$, we can choose\footnote{Notice that this transformation is inside a $SU(2)$ subgroup of $SU_R(4)$ and keeps ${\mathrm{diag}}(-i, -i, i, i)$ invariant.  So its action on the Wilson loop only changes $\bar{\alpha}^I, \beta_I$. }
\be \bar{\alpha}^I=(\bar{\alpha}, 0, 0, 0)\,,\,\quad
\beta_I=(\beta, 0, 0, 0)\,,\label{alphabeta}\ee
with the constrains $\bar{\alpha}\beta=i$. Now $U_{I}^{\,\,\,J}=\mathrm{diag}(i, -i, -i, -i)$, $\bar{f}_1$ and $f_2$ become,
\be\bar{f}_1=\sqrt{\frac{2\pi}{k}}\bar{\alpha}\bar{\zeta}\psi_1|\dot{x}|\,,\,\quad
f_2=\sqrt{\frac{2\pi}{k}}\psi^{\dagger 1}\eta\beta|\dot{x}|\,.\ee
The equivalence relation~(\ref{equivalence}) now becomes,
\be (\bar{\alpha}, \beta)\sim (\lambda \bar{\alpha}, \lambda^{-1}\beta)\,, \, \lambda\in \mathbf{C}^\ast\,. \ee
We will denote this half-BPS Wilson loops by
$W^{1+}_{1/2}$ and the corresponding super-connection by $L^{1+}_{1/2}$. The supersymmetries preserved by $W^{1+}_{1/2}$ are,
\be \vartheta_{1I}=iR^{-1}\gamma_3\theta_{1I}\,,\,\quad \epsilon^{1IJK}\vartheta_{JK}=-i R^{-1} \gamma_3 \epsilon^{1IJK}\theta_{JK}\,. \ee

\paragraph{$1/3$-BPS circular WLs.}
In the construction of $1/3$-BPS Wilson loops, we start with the following super-connection $L^{4-}_{1/2}$ in another half-BPS Wilson loop $W^{4-}_{1/2}$,
\bea  L^{4-}_{1/2}=\left(\begin{array}{cc}
                           \mathcal{A}& \bar{f}_1\\
                           f_2&\hat{\mathcal{A}}\end{array}\right)\,, \eea
with
\bea
&& \mathcal{A}=A_\mu \dot{x}^\mu+\frac{2\pi}{k}|\dot{x}|\tilde{U}_I^{\,\,\,J} Y^I Y^\dagger_J\,,\,\,\quad
\bar{f}_1=\frac{2\pi}{k}\bar{\rho}\bar{\mu}\psi_4|\dot{x}|\,,\nonumber\\
&&\hat{\mathcal{A}}=\hat{A}_\mu\dot{x}^\mu+\frac{2\pi}{k}|\dot{x}|\tilde{U}_I^{\,\,\,J} Y^\dagger_J Y^I\,,\,\,\quad
f_2=\frac{2\pi }{k}\psi^{\dagger4}\nu\delta |\dot{x}|\,,\nonumber\\
&&\tilde{U}_I^{\,\,\,J}=\left(\begin{array}{cccc}
                              i & 0 & 0 &0  \\
                              0 & i & 0 &0  \\
                             0  & 0 & i & 0 \\
                               0& 0 &0 & -i
                            \end{array}\right)\,, \nonumber\\
&& \bar{\mu}^\alpha=(e^{i\tau/2}, -e^{-i\tau/2})\,,\,\quad  \nu_\alpha=\left(-e^{-i\tau/2}, e^{i\tau/2}\right) \,.                         \eea
Here $\bar{\rho}, \delta$ are two complex numbers satisfying $\bar{\rho}\delta=-i$,  and we have the equivalence relation
\be (\bar{\rho}, \delta)\sim (\lambda \bar{\rho}, \lambda^{-1}\delta)\,, \, \lambda\in\mathbf{C}^\ast\,. \ee
The corresponding Wilson loop
\be W^{4-}_{1/2}=\mathrm{Tr} \mathcal{P} \, \exp \left(-i\oint d\tau L^{4-}_{1/2}(\tau)\right)\,, \ee
preserve the following supersymmetries,
\be \vartheta_{I4}=-i\gamma_3 R^{-1}\theta_{I4}\,, \, \epsilon^{4IJK}\vartheta_{JK}=i\gamma_3R^{-1}\epsilon_{4IJK}\vartheta_{JK}. \ee
This Wilson loop belong to the class II of the fermionic $1/6$-BPS Wilson loops in~\cite{Ouyang:2015iza, Ouyang:2015bmy}.

The $1/3$-BPS Wilson loops are constructed by $L^{1+}_{1/2}$ and $L^{4-}_{1/2}$,
\bea  W_{1/3}=\mathrm{Tr} \mathcal{P} \, \exp \left(-i\oint d\tau L_{1/3}(\tau)\right)\,,\nonumber\\
L_{1/3}={\mathrm{diag}}(\underbrace{L_{1/2}^{1+}, \cdots, L_{1/2}^{1+}}_{n_1},\underbrace{ L^{1/2}_{4-}, \cdots, L_{1/2}^{4-}}_{n_4})\,.
\label{onethird}\eea
Notice that $L_{1/3}$ is a $((n_1+n_4)N|(n_1+n_4)N)$ supermatrix, since both $L^{1+}_{1/2}$ and $L^{4-}_{1/2}$ are $(N|N)$ supermatrices.

The supersymmetries preserved by $W_{1/3}$ are given by the following ones shared by $L^{1+}_{1/2}$ and $L^{4-}_{1/2}$,
\bea  &&\vartheta_{12}=i\gamma_3\theta_{12}\,, \, \vartheta_{13}=i\gamma_3\theta_{13}\,, \, \vartheta_{24}=-i\gamma_3\theta_{24}\,,\, \vartheta_{34}=-i\gamma_3\theta_{34}\,, \\
&&\theta_{14}=\theta_{23}=\vartheta_{14}=\vartheta_{23}=0\,.\eea

\section{Wilson loop one-point function in ABJM theory}
\label{sc:onepoint}
\subsection{The boundary states from  Wilson loops}
\label{ssc:boundary}
The main goal of this paper is to study the tree-level correlation function of the BPS Wilson loops reviewed in the previous section and the single-trace operator, \be \mathcal{O}_C=C^{J_1\cdots J_L}_{I_1\cdots I_L} \mathrm{Tr}(Y^{I_1}Y^\dagger_{J_1}\cdots Y^{I_L}Y^\dagger_{J_L})\,,\ee
in the scalar sector. The coefficients  $C^{J_1\cdots J_L}_{I_1\cdots I_L}$ are chosen such that this single-trace operator is the eigenstate of the planar two-loop dilatation operator. The single-trace  operator is put at the origin of the three-dimensional space. The Wilson loops considered in this paper are in the fundamental representation of a suitable  (super-)group. More precisely, the bosonic  $1/6$-BPS Wilson loop~(\ref{bosonic}) is  in the fundamental representation of $U(N)$. $W^{B, \mathrm{big}}_{1/6}$ is in the fundamental representation of the gauge group $U(N)\times U(N)$. The fermionic $1/6$-BPS Wilson loop~(\ref{fermionic16}) and the half-BPS Wilson one
are  in the fundamental representation of the supergroup $U(N|N)$. Finally,  the $1/3$-BPS Wilson loop~(\ref{onethird}) is  in the fundamental representation of the supergroup $U((n_1+n_4)N|(n_1+n_4)N)$.

\begin{figure}[h!]
\centering
\includegraphics[scale=0.4]{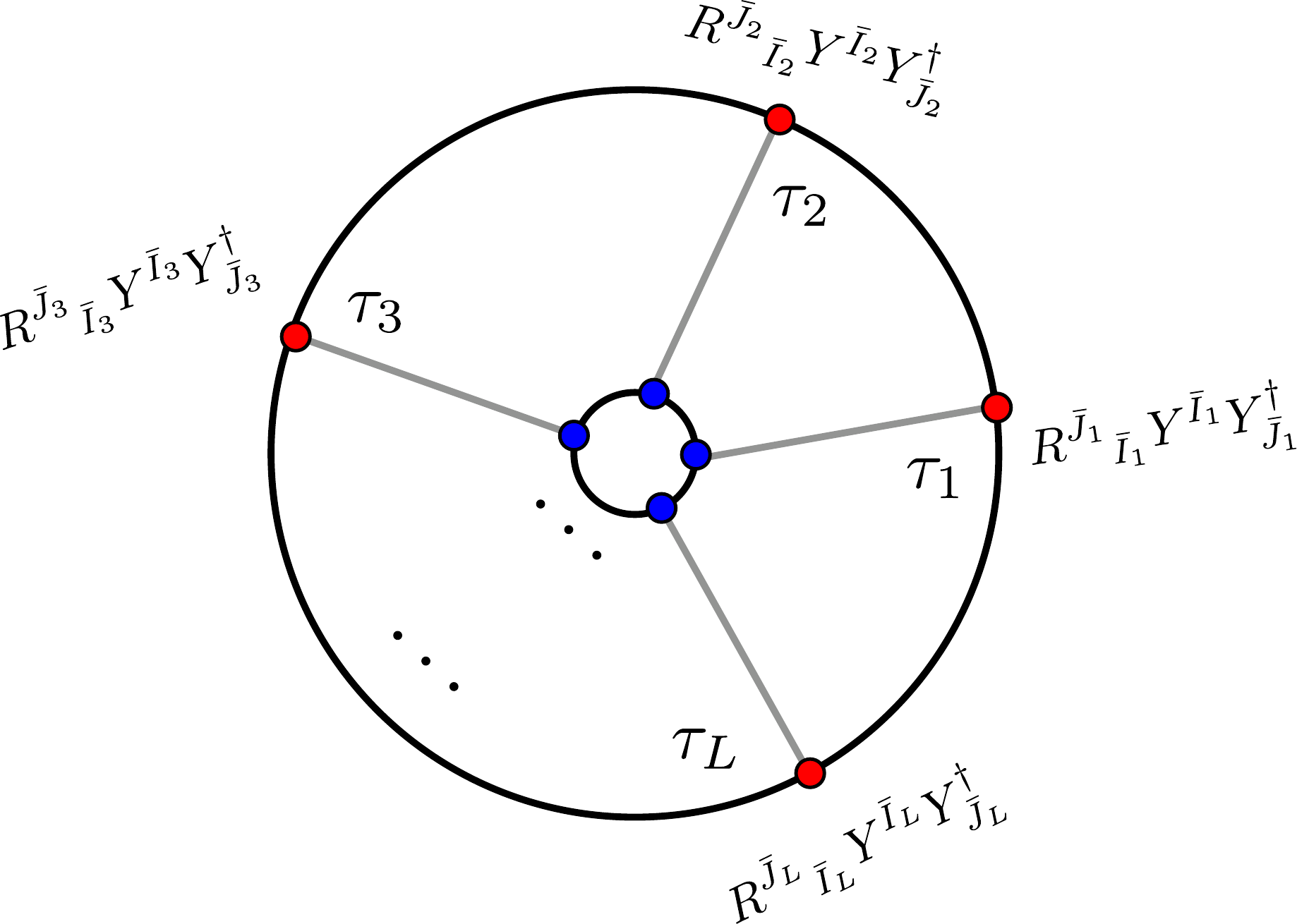}
\caption{Planar Wick contractions between the local operator and the Wilson loop. {Here each blue or red dot indicates  a pair of scalar fields and each straight line indicates a pair of contractions. The outer circle indicates the Wilson loop, and the inner circle indicates the single-trace operator. Although the single-trace operator is local in space-time, we draw it as a circle to make the color structure of the operator and the contraction more clear. }}
\label{fig:WL}
\end{figure}

At tree level, the correlator $\langle W(\mathcal{C})^B_{1/6} \mathcal{O}_C(0)\rangle$ only gets contributions from
\bea  &&\oint \cdots \oint d\tau_{1>2>\cdots >L}\left(\frac{2\pi}{k}\right)^L \langle
{\mathrm{tr}}(R^{\tilde{J}_1}_{\,\,\,\,\,\tilde{I}_1} Y^{\tilde{I}_1}(x_1) Y^\dagger_{\tilde{J}_1}(x_1)\cdots R^{\tilde{J}_L}_{\,\,\,\,\,\tilde{I}_L} Y^{\tilde{I}_L}(x_L) Y^\dagger_{\tilde{J}_L}(x_L))\nonumber\\
&& C^{J_1\cdots J_L}_{I_1\cdots I_L} {\mathrm{tr}}(Y^{I_1}(0)Y^\dagger_{J_1}(0)\cdots Y^{I_L}(0)Y^\dagger_{J_L}(0))
\rangle\,,  \eea
where $x_i=(R\cos\tau_i, R\sin\tau_i, 0), \, i=1, \cdots, L$, and
\be \oint \cdots \oint d\tau_{1>2>\cdots >L} =\int_0^{2\pi}d\tau_1\int_0^{\tau_1}d\tau_2\cdots \int_0^{\tau_{L-1}}d\tau_L\,. \ee
In the large $N$ limit, we only take into account planar Wick contractions, as is shown in figure~\ref{fig:WL}.
One can easily obtain
\be \langle W(\mathcal{C})^B_{1/6} \mathcal{O}_C(0)\rangle=\frac{\lambda^{2L}k^L}{(L-1)!(2R)^{2L}}
 C^{J_1\cdots J_L}_{I_1\cdots I_L}  R^{I_L}_{\,\,\,\,J_L}\cdots R^{I_1}_{\,\,\,\,J_1}\,,\ee
 where $\lambda\equiv\frac{N}{k}$ is the 't~Hooft coupling of ABJM theory and the tree-level propagators of the scalar fields~(\ref{propagator})  have been used.
For later convenience,
we introduce the following two-site boundary state which is specified by a $4\times 4$ matrix ${M}$ as
\bea
\label{boundaryState} \langle\mathcal{B}_M|&\equiv& M^{I_1}_{\,\,\,\,J_1}M^{I_ 2}_{\,\,\,\,J_2}\cdots M^{I_L}_{\,\,\,\,J_L}\langle I_1, J_1, \cdots, I_L, J_L|\nonumber\\
&=&\left(M^I_{\,\,J}\langle I, J|\right)^{\otimes L}\,. \eea
Then \be |\mathcal{B}_M\rangle=\left((M^I_{\,\,J})^\ast|I, J\rangle\right)^{\otimes L} \ee
Our convention for the Hermitian conjugation of  the spin chain states
is
\bea  && \left(\langle I_1, J_1, \cdots, I_L, J_L |\right)^\dagger=|I_1, J_1\cdots, I_L, J_L \rangle\,.
\eea
We define the 1/6-BPS boundary state as\footnote{Notice that ${R^I}_J$ should not be confused with $R$ which is the radius of the circular Wilson loop.}
    \be\label{boundary16} |\mathcal{B}_{1/6}^B\rangle=|\mathcal{B}_R\rangle\,,
\ee
where ${R^I}_J=\text{diag}(i,i,-i,-i)$.  Then the above correlation function can be expressed as
\be  \langle W(\mathcal{C})^B_{1/6} \mathcal{O}_C(0)\rangle=\frac{\lambda^{2L}k^L}{(L-1)!(2R)^{2L}} \langle \mathcal{B}_{1/6}^B |\mathcal{O}_C\rangle\,,\ee
where $|\mathcal{O}_C\rangle{\equiv C^{J_1\cdots J_L}_{I_1\cdots I_L}|I_1, J_1, \cdots, I_L, J_L\rangle}$ is the spin chain state corresponding to the operator $\mathcal{O}_C$. 
Our convention for the overlap of  two spin chain states
is
\bea 
&&\langle  I_1, J_1, \cdots, I_L, J_L  |   M_1, N_1, \cdots, M_L, N_L\rangle=\delta_{I_1}^{M_1}\delta^{J_1}_{N_1}\cdots \delta_{I_L}^{M_L}\delta^{J_L}_{N_L}\,.
\eea
Let us define the normalization factor $\mathcal{N}_{\mathcal{O}}$ using the two-point function of $\mathcal{O}$
and $\mathcal{O}^\dagger$ as
\be  \langle\mathcal{O}(x)\mathcal{O}^\dagger(y)\rangle=\frac{\mathcal{N}_{\mathcal{O}}}{|x-y|^{2\Delta_{\mathcal{O}}}}\,,\ee
where $\Delta_{\mathcal{O}}$ is the conformal dimension of $\mathcal{O}$.
At tree level and  the planar limit, we have
\be \mathcal{N}_{\mathcal{O}}=\left(\frac{N}{4\pi}\right)^{2L}L\langle\mathcal{O} | \mathcal{O}\rangle\,.\ee
We define the Wilson-loop one-point function
as
\be\langle\!\langle \mathcal{O}\rangle\! \rangle_{W(\mathcal{C})}\equiv \frac{\langle W(\mathcal{C}) \mathcal{O} \rangle}{\sqrt{\mathcal{N}_\mathcal{O}}}\,.\ee
Then for $W^B_{1/6}$ we have
\be
\label{eq:WB16}
\langle\!\langle \mathcal{O}\rangle\!\rangle_{W(\mathcal{C})^B_{1/6}}
=\frac{\pi^L\lambda^L }{R^{2L}(L-1)!\sqrt{L}}\frac{\langle \mathcal{B}_{1/6}^
B|\mathcal{O} \rangle}{\sqrt{\langle \mathcal{O} |\mathcal{O}\rangle}}\,.
\ee
The computation of the Wilson loop one-point function thus amounts to the calculation of \be
\label{eq:overDef}
\frac{\langle \mathcal{B}_{1/6}^
B|\mathcal{O} \rangle}{\sqrt{\langle \mathcal{O} |\mathcal{O}\rangle}}\,.\ee

Similar computations for other Wilson loops studied in this paper can also be reduced to the computation of overlaps of the form in \eqref{eq:overDef}, with the corresponding boundary states.
For $\hat{W}(\mathcal{C})^B_{1/6}$,  the boundary state is
\begin{equation}
\label{eq:BWh16}
\langle\hat{\mathcal{B}}^B_{1/6}|=R^{I_1}_{\,\,\,\,J_L}R^{I_2}_{\,\,\,\,J_1}\cdots R^{I_L}_{\,\,\, \, J_{L-1}}\langle I_1, J_1, \cdots, I_L, J_L|\,.
\end{equation}
We can rewrite $|\hat{\mathcal{B}}_{1/6}^B\rangle$ as
\be |\hat{\mathcal{B}}_{1/6}^B\rangle=U_{\rm even}|\mathcal{B}_{1/6}^B\rangle\,\ee
 where $U_{\rm even}$ is the shift operator which shifts all even site to the left by two units and leave the odd sites untouched,
 \be\label{Ueven}  U_{\rm even}|I_1, J_1, I_2, J_2, \cdots, I_{L-1}, J_{L-1}, I_L, J_L\rangle=|I_1, J_2, I_2, J_3, \cdots, I_{L-1}, J_L, I_L, J_1\rangle.\ee
{Combining \eqref{boundary16} and \eqref{eq:BWh16}, we obtain the boundary state of $W(\mathcal{C})^{B, \mathrm{big}}_{1/6}$
\bea
\label{eq:BBig16}
|\mathcal{B}_{1/6}^{B, \mathrm{big}}\rangle=|\mathcal{B}_{1/6}^{B}\rangle+|\hat{\mathcal{B}}_{1/6}^{B}\rangle
=(1+U_{\rm even})|\mathcal{B}_R\rangle\,.
\eea
For $W(\mathcal{C})^F_{1/6}$, we simply replace ${R^I}_J$ by ${U^I}_J$ given in~(\ref{fconstruction}),
\be | \mathcal{B}_{1/6}^F \rangle=(1+U_{\rm even})|\mathcal{B}_U\rangle \,. \ee
The boundary state  corresponding to  $W(C)_{1/2}$}, which will be denoted by $| \mathcal{B}_{1/2} \rangle$ is given by $| \mathcal{B}_{1/6}^F\rangle$ with the additional constraints~\eqref{constraints}.

{In particular, the boundary state $|\mathcal{B}_{1/2}^{1+}\rangle$ corresponding to $W(C)_{1/2}^{1+}$ is
\be |\mathcal{B}_{1/2}^{1+}\rangle=(1+U_{\rm even})|\mathcal{B}_U\rangle\,,  \ee
 with ${U^I}_J=\mathrm{diag}(i, -i, -i, -i)$. Similarly,
the boundary state $|\mathcal{B}_{1/2}^{4-}\rangle$ corresponding to $W(C)_{1/2}^{4-}$ is
\be |\mathcal{B}_{1/2}^{4-}\rangle=(1+U_{\rm even})|\mathcal{B}_{\tilde{U}}\rangle\,,  \ee
with ${\tilde{U}^I}_{\,\,J}=\mathrm{diag}(i, i, i, -i)$.

Finally, for $W(\mathcal{C})_{1/3}$, we have
\be | \mathcal{B}_{1/3}\rangle=n_1|\mathcal{B}_{1/2}^{1+}\rangle+n_4|\mathcal{B}_{1/2}^{4-}\rangle\,. \ee}

\subsection{Integrable and non-integrable boundary states}
\label{ssc:integrability}
In this subsection, we will prove that the boundary states $|\mathcal{B}_{1/6}^B\rangle, |\hat{\mathcal{B}}_{1/6}^B\rangle, | \mathcal{B}_{1/2}\rangle$ and $|\mathcal{B}_{1/3}\rangle$
are integrable  at tree level by employing the method proposed in~\cite{Piroli:2017sei}.  We will also show that the $| \mathcal{B}_{1/6}^F\rangle$ with $\bar{\alpha}^2=\beta_2=0$ is not integrable unless $\bar{\alpha}^1\beta_1=0,i$.\par

Let us consider the  boundary state
$|\mathcal{B}_M\rangle$ defined by a matrix $M$ as in~(\ref{boundaryState}). In what follows, we will encounter several examples in which $M$ is a diagonal matrix.
In this case, the overlap $\langle \mathcal{B}_M  |  \mathbf{u}, \mathbf{v}, \mathbf{w}\rangle$ is nonzero only if
the numbers of the Bethe roots, $K_{\mathbf{u}}, K_{\mathbf{v}}, K_{\mathbf{w}}$, and the length of the spin chain $2L$, satisfy $K_{\mathbf{u}}=K_{\mathbf{v}}=K_{\mathbf{w}}=L$. Notice that this selection rule has nothing to do with integrability of the boundary state.

 In the algebraic Behte ansatz approach, the $SU(4)$ sector $R$-matrices of the ABJM theory at two-loop level are given by
\begin{align}
\begin{array}{l}
R_{12}^{\bullet \bullet}(u)=R_{12}^{\circ \circ}(u)=u+P_{12}\equiv R_{12}(u)\,, \\
R_{12}^{\bullet \circ}(u)=R_{12}^{\circ \bullet}(u)=-u-2+K_{12}\equiv\bar{R}_{12}(u) \,,
\end{array}
\end{align}
where $\bullet$ denotes the states in the $\mathbf{4}$ representation of $SU(4)_R$, while $\circ$ denotes the states in the $\bar{\mathbf{4}}$ representation.
{The $R$-matrices satisfy the following crossing symmetry relations
\begin{align}
R_{12}\label{crossingSymmetry}(u)^{t_{1}}=\bar{R}_{12}(-u-2),\qquad \bar{R}_{12}(u)^{t_{1}}=R_{12}(-u-2) \,,
\end{align}
 and the relation
\bea  R_{12}(u)^{t_{1}t_2}&=&R_{12}(u),\qquad \bar{R}_{12}(u)^{t_{1}t_2}=\bar{R}_{12}(u),\nonumber\\
 P_{12}R_{12}(u)P_{12}&=&R_{12}(u), \qquad P_{12}\bar{R}_{12}(u)P_{12}=\bar{R}_{12}(u)\,,\eea
where $t_i$ denotes transposition in the $i$-th space. One key feature of algebraic Bethe ansatz approach of the ABJM spin chain is that it requires two $R$-matrices, $R(u)$ and $\bar{R}(u)$, due to its alternating nature. This is different from the case when the spin on each site is in the same representation.}

In the following, we will show that when there exists a four-dimensional matrix $K(u)$ satisfying the boundary Yang-Baxter equation (BYBE)\footnote{{Notice that here we only need to use the BYBE involving one of the $R$-matrices, $R(u)$.  }}
\begin{align}\label{BYBE}
R_{12}(u-v) K_{1}(u) R_{12}(u+v) K_{2}(v)=K_{2}(v) R_{12}(u+v) K_{1}(u) R_{12}(u-v)\,,
\end{align}
 then the boundary state $|\mathcal{B}_M\rangle$ with $M=K(-1)^\ast$ is integrable in the sense that it satisfies the following twisted integrability condition~\cite{Gombor:2020kgu, Gombor:2022deb},
\begin{align}\label{integrablboundary}
	\tau(-u-2)|\mathcal{B}_M\rangle=\tau(u)|\mathcal{B}_M\rangle\,,
\end{align}
or equivalently~\cite{Yang:2022dlk},
$\Pi \bar{\tau}(u) \Pi|\mathcal{B}_M\rangle=\tau(u)|\mathcal{B}_M\rangle$\,,
where
\begin{align}\label{tau}
	\tau(u) & =\operatorname{Tr}_{0}\left(R_{01}(u) \bar{R}_{02}(u) \cdots R_{0, 2L-1}(u) \bar{R}_{0, 2L}(u)\right)\,, \\
	\bar{\tau}(u) & =\operatorname{Tr}_{0^\prime}\left(\bar{R}_{0^\prime 1}(u) R_{0^\prime 2}(u) \cdots \bar{R}_{0^\prime, 2L-1}(u) R_{0^\prime, 2L}(u)\right)\,,
\end{align}
are the transfer matrices. Here $0, 0^\prime$ denote two auxiliary spaces and $\Pi$
is the parity operator
\be   \Pi|I_1, J_1, \cdots, J_{2L}\rangle = |J_{2L}, \cdots, J_2, I_1\rangle\,.\ee
Using the explicit forms of eigenvalues of $\tau(u)$ and $\bar{\tau}(u)$~\cite{Yang:2022dlk}, we conclude that for integrable boundary states, the overlap
$\langle \mathcal{B}_M  |  \mathbf{u}, \mathbf{v}, \mathbf{w}\rangle$ is non-zero only if the selection rules
\begin{align}
\label{selectionRules}
\mathbf{u}=-\mathbf{v}\,,\qquad \mathbf{w}=-\mathbf{w}\,
\end{align}
are satisfied.\par

\begin{figure}[h!]
    \centering
\includegraphics[scale=0.3]{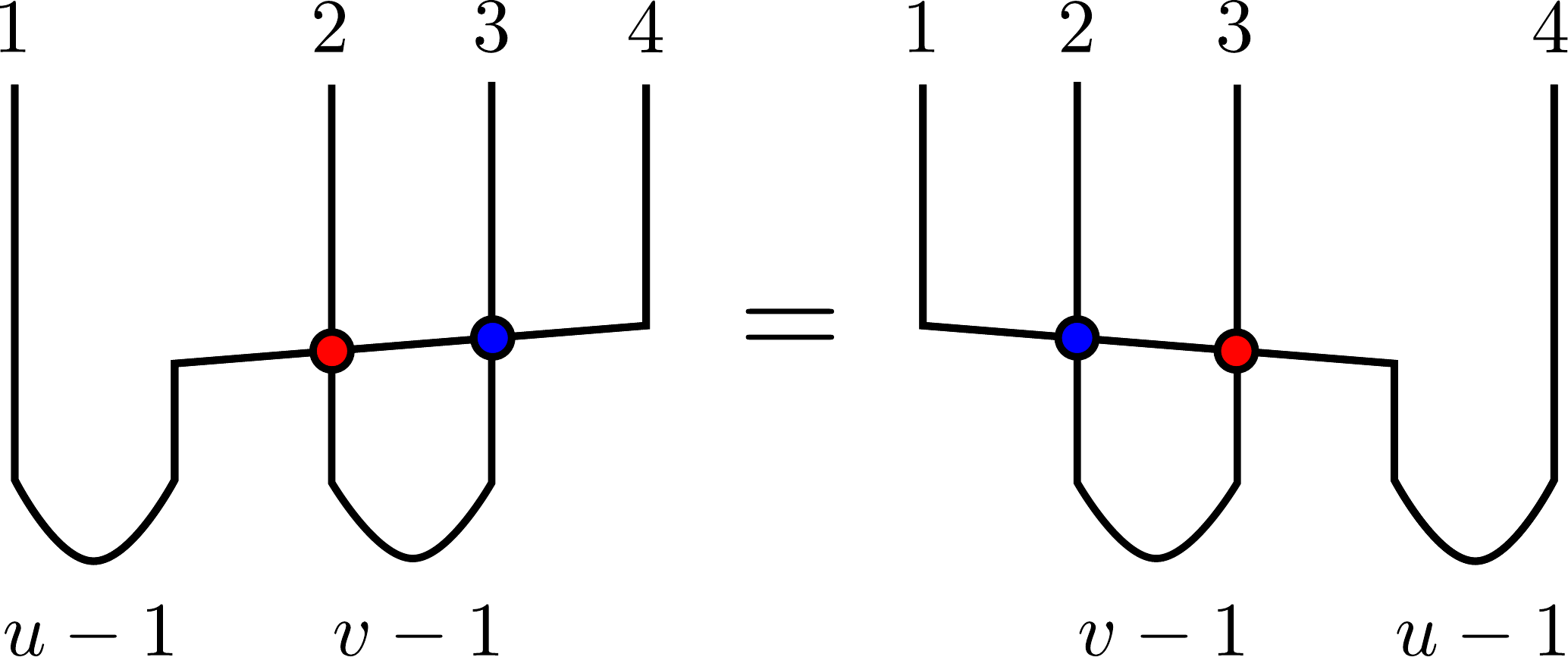}
    \caption{A pictorial representation of eq.~(\ref{eq:RR}) from the boundary Yang-Baxter equation. Here the red node denotes $\check{\bar{R}}(-u-v)$ and the blue node denotes $\check{R}(v-u)$.}
    \label{fig:RR}
\end{figure}

Let us define the state
\begin{align}
|\phi(u-1)\rangle_{a b}=K_{J}^{I}(u-1)|I\rangle_{a} \otimes|J\rangle_{b}\,.
\end{align}
The boundary Yang-Baxter equation (\ref{BYBE}) leads to
\begin{align}
\begin{aligned}\label{eq:RR}
& \check{R}_{34}(v-u) \check{\bar{R}}_{23}(-u-v)\left|\phi_{0}(u-1)\right\rangle_{12} \otimes\left|\phi_{0}(v-1)\right\rangle_{34} \\
= & \check{R}_{12}(v-u) \check{\bar{R}}_{23}(-u-v)\left|\phi_{0}(v-1)\right\rangle_{12} \otimes\left|\phi_{0}(u-1)\right\rangle_{34}\,,
\end{aligned}
\end{align}
where  $\check{R}_{12}=P_{12} R_{12}$ and $\check{\bar{R}}_{12}=P_{12} \bar{R}_{12}$. This relation can be shown pictorially as in figure~\ref{fig:RR}.

{Notice that $\bar{R}(u)$ also appears here, since we have used one of the crossing symmetry relations~(\ref{crossingSymmetry}).}

Introducing two four-dimensional  auxiliary spaces  $h_{0}$  and  $h_{2L+1}$ and
following the derivation in the Section~4 and Appendix~C of~\cite{Piroli:2017sei}, we can prove that
\begin{align}
\label{eq:longR}
\begin{aligned}
& \check{R}_{2 L+1,2 L}(v-u) \check{\bar{R}}_{2 L, 2 L-1}(-v-u) \cdots \check{R}_{32}(v-u) \check{\bar{R}}_{21}(-v-u) \\
& |\phi(u-1)\rangle_{01}\otimes|\phi(v-1)\rangle_{23} \otimes\cdots\otimes|\phi(v-1)\rangle_{2 L, 2 L+1} \\
= & \check{R}_{2 L+1,2 L}(v-u) \check{\bar{R}}_{2 L, 2 L-1}(-v-u) \cdots \check{R}_{01}(v-u) \check{\bar{R}}_{12}(-v-u) \\
& |\phi(v-1)\rangle_{01}\otimes|\phi(u-1)\rangle_{23}\otimes |\phi(v-1)\rangle_{45} \otimes\cdots\otimes|\phi(v-1)\rangle_{2 L, 2 L+1} \\
= & \cdots\\
= & \check{R}_{01}(v-u) \check{\bar{R}}_{12}(-v-u) \cdots \check{R}_{2 L-2,2 L-1}(v-u) \check{\bar{R}}_{2 L-1,2 L}(-v-u) \\
& |\phi(v-1)\rangle_{01}\otimes|\phi(v-1)\rangle_{23}\otimes \cdots\otimes|\phi(v-1)\rangle_{2 L-2,2 L-1}\otimes|\phi(u-1)\rangle_{2 L, 2 L+1}\,.
\end{aligned}
\end{align}

\begin{figure}
    \centering  \includegraphics[scale=0.3]{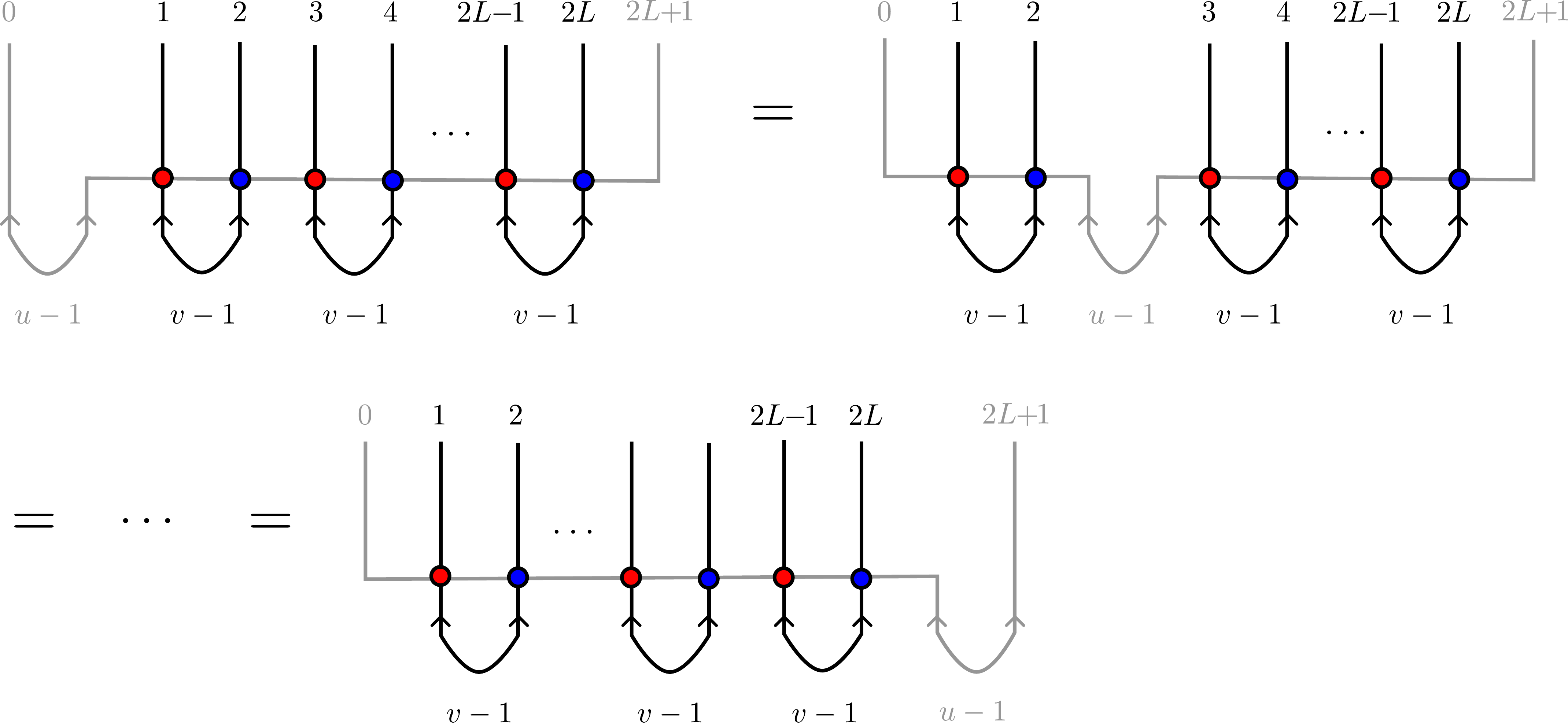}
    \caption{A pictorial derivation of \eqref{eq:longR}.}
    \label{fig:longR}
\end{figure}

This in turn implies
\begin{align}\label{finalstate}
\begin{aligned}
&\operatorname{Tr}_{0}\left(R_{01}(-u-2) \bar{R}_{02}(-u-2) \cdots R_{0,2 L-1}(-u-2) \bar{R}_{0,2 L}(-u-2)\right)|\mathcal{B}_M\rangle \\
=&\operatorname{Tr}_{0}\left(R_{01}(u) \bar{R}_{02}(u) \cdots R_{0,2 L-1}(u) \bar{R}_{0,2 L}(u)\right)|\mathcal{B}_M\rangle\,.
\end{aligned}
\end{align}
Here the condition  $K(-1)=M^\ast$ has been used. Pictorially the above derivation is shown in figure~\ref{fig:longR}.

In terms of the transfer matrices $\tau(u)$ and $\bar{\tau}(u)$ given in (\ref{tau}),
\eqref{finalstate} can be written as
\begin{align}	\tau(-u-2)|\mathcal{B}_M\rangle=\tau(u)|\mathcal{B}_M\rangle.
\end{align}
As mentioned above,  this equation  is equivalent~\cite{Yang:2022dlk} to the twisted integrability condition~\cite{Gombor:2020kgu, Gombor:2022deb},  $\Pi \bar{\tau}(u) \Pi|\mathcal{B}_M\rangle=\tau(u)|\mathcal{B}_M\rangle$ \footnote{In fact, there exist untwisted integrability condition~\cite{Gombor:2020kgu, Gombor:2022deb} $\Pi \tau(u) \Pi|\mathcal{B}_M\rangle=\tau(u)|\mathcal{B}_M\rangle, \Pi \bar{\tau}(u) \Pi|\mathcal{B}_M\rangle=\bar{\tau}(u)|\mathcal{B}_M\rangle$, which  leads to the selection rule $\mathbf{u}=-\mathbf{u}, \mathbf{v}=-\mathbf{v}, \mathbf{w}=-\mathbf{w}$. But such case has not appeared in the boundary states in the ABJM theory yet~\cite{Yang:2021hrl, Kristjansen:2021abc}.}.
This finishes the proof about the integrability of the boundary state $|\mathcal{B}_M\rangle$ assuming that there exists the matrix $K(u)$ satisfying the BYBE,  (\ref{BYBE}), and $K(-1)=M^\ast$.
We can similarly prove that for such $M$, the boundary state $|\hat{\mathcal{B}}_M\rangle\equiv U_{\rm even }|{B}_M\rangle$ is also integrable and leads to the same selection rule\footnote{This result can be also obtained from the general classification of integrable boundary states in~\cite{Gombor:2021hmj}.}.

{ Now we turn to boundary states from some BPS Wilson loops we list above. For  $|\mathcal{B}_{1/6}^B\rangle(=|\mathcal{B}_{R}\rangle)$ and  $|\hat{\mathcal{B}}_{1/6}^B\rangle(=|\hat{\mathcal{B}}_R\rangle)$, one can check that the matrix $K(u)=R^\ast$ is the solution of BYBE (\ref{BYBE}), so these two boundary states are both  tree-level integrable. Then $| \mathcal{B}_{1/6}^{B, \mathrm{big}}\rangle$ is   tree-level integrable as well. As for the boundary state $|\mathcal{B}_{1/2}^{1+}\rangle(=(1+U_{\rm even})|\mathcal{B}_U\rangle)$,  we can just choose $K(u)=U^\ast=\mathrm{diag}(-i, i, i, i)$ since it also satisfies the BYBE  (\ref{BYBE}). Then we get that  $|\mathcal{B}_{1/2}^{1+}\rangle$ is integrable at tree level. Since the tree-level integrability of $|\mathcal{B}_M\rangle$ is preserved when we perform  a $SU(4)_R$ transformation on $M$ or multiply $M$ by a constant, we get that the generic $|\mathcal{B}_{1/2}\rangle$ and $|\mathcal{B}_{1/2}^{4-}\rangle$ are also tree-level integrable.
This leads to the conclusion that $|\mathcal{B}_{1/3}\rangle$ is tree-level integrable.

Now we turn consider the boundary state
\bea
|\mathcal{B}_{1/6}^F \rangle&=&|\mathcal{B}_U \rangle+| \hat{\mathcal{B}}_U \rangle
=(1+U_{\rm even})|\mathcal{B}_U\rangle\,,
\eea
with $U=\mathrm{diag}(i, i-2\epsilon, -i, -i)$. This boundary state
corresponds to the generic fermionic $1/6$-BPS Wilson loops with $\bar{\alpha}^2=\beta_2=0$ and $\epsilon$ is given by $\epsilon= \bar{\alpha}^1\beta_1$.}

 In the following we will show that when $\epsilon\neq 0, i$, this state is not integrable. The idea is to employ the following set of Bethe roots with $L=3, K_{\mathbf{u}}=K_{\mathbf{w}}=1, K_{\mathbf{v}}=2$,
\be
u_1=0.866025,\,\quad w_1= 0.866025,\,\quad v_1=-0.198072,\,\quad v_2= 0.631084\,,
\ee
which does not satisfy the selection rule $\mathbf{u}=-\mathbf{v}, \mathbf{w}=-\mathbf{w}$. Notice that the set of roots also satisfy the zero momentum condition.
However these roots do not satisfy the first selection rule $K_{\mathbf{u}}=K_{\mathbf{v}}=K_{\mathbf{w}}=L$. This fact leads to the result that the overlap $\langle\mathcal{B}_{1/6}^F|\mathbf{u}, \mathbf{v}, \mathbf{w}\rangle=0$ for this set of Bethe roots and whether the $|\mathcal{B}_{1/6}^F\rangle$ is  integrable or not can not be detected
by this result.
The way out is to perform the following $SO(4)\subset SU(4)_R$ transformation~\cite{Gombor:2022aqj},
\be  U_\theta=g(\theta)  U g(\theta)^{-1}\,,\ee
with
\begin{align}
g(\theta)=\left(
\begin{array}{cccc}
\cos ^2\theta & \sin \theta & 0 & \sin \theta \cos \theta \\
-\sin \theta \cos ^2\theta & \cos ^2\theta & \sin \theta & -\sin ^2\theta \cos \theta \\
\sin ^2\theta \cos \theta & -\sin \theta \cos \theta & \cos \theta & \sin ^3\theta \\
-\sin \theta & 0 & 0 & \cos \theta \\
\end{array}
\right)\,,\label{equation:oldg}
\end{align}
where $\theta$ satisfies  $0<\theta<\frac{\pi}{2}$.
Due to $SU(4)_R$ invariance,  $|\mathcal{B}_{1/6}^F\rangle$ is integrable if and only if   $|\mathcal{B}_{1/6, \,\theta}^F\rangle\equiv (1+U_{\rm even})|\mathcal{B}_{U_\theta}\rangle$ is.
Through direct computation, we found that $\langle \mathcal{B}_{1/6,\, \theta}^F| \bf{u}, \bf{v}, \bf{w}\rangle$ is zero if and only if $\epsilon=0$ or $\epsilon=i$.
This shows that generic fermionic $1/6$-BPS Wilson loop gives non-integrable boundary state. More precisely, such boundary state satisfies neither the twisted condition nor the untwisted one.  Notice that, when $\epsilon=i$, there is supersymmetric enhancement for this fermionic $1/6$-BPS Wilson loop and it in fact becomes half-BPS. When $\epsilon=0$, the bosonic part of this fermionic $1/6$-BPS Wilson loop is the same as the big bosonic $1/6$-BPS Wilson loop $W_{1/6}^{B, \mathrm{big}}$.\footnote{If we consider the vacuum expectation value of the Wilson loop or the correlators of the Wilson loop with operators out of it, this fermionic $1/6$-BPS Wilson loop with $\epsilon=0$ is identidical  to the bosonic $1/6$-BPS Wilson loop $W_{1/6}^{B, \mathrm{big}}$~\cite{Drukker:2020opf}.} We have already shown that $\epsilon=0$ and $\epsilon=i$ lead to integrable boundary states.

To sum up, we find that only some of the supersymmetric Wilson loops correspond to tree-level integrable boundary states, they are listed as follows
\begin{itemize}
\item The bosonic $1/6$-BPS Wilson loop corresponding to the state
\begin{align}
\label{eq:BB16}
|\mathcal{B}_{1/6}^{B, \mathrm{big}}\rangle=(1+U_{even})|\mathcal{B}_R\rangle,
\qquad {R^I}_J=\text{diag}(i,i,-i,-i)
\end{align}
\item The $1/2$-BPS Wilson loop corresponding to the state
\begin{align}
\label{eq:halfBPSp}
|\mathcal{B}_{1/2}^{1+}\rangle=(1+U_{\text{even}})|\mathcal{B}_U\rangle,\qquad {U^I}_J=\text{diag}(i,-i,-i,-i)
\end{align}
\item The $1/2$-BPS Wilson loop corresponding to the state
\begin{align}
\label{eq:halfBPSm}
|\mathcal{B}_{1/2}^{4-}\rangle=(1+U_{\text{even}})|\mathcal{B}_{\tilde{U}}\rangle,\qquad{{\tilde{U}}^I}_{\phantom{I}J}=\text{diag}(i,i,i,-i)\,.
\end{align}
\item The $1/3$-BPS Wilson loop corresponding to the state
\begin{align}
|\mathcal{B}_{1/3}\rangle=n_1|\mathcal{B}_{1/2}^{1+}\rangle+n_4|\mathcal{B}_{1/2}^{4-}\rangle
\end{align}
\end{itemize}
Notice that in the above states, both $|\mathcal{B}_M\rangle$ and $U_{\text{even}}|\mathcal{B}_M\rangle$ ($M=R,U,\tilde{U}$) are integrable at tree level.
In the next section, we will derive the exact overlap formula of the tree-level integrable boundary states and the on-shell Bethe states of the SU(4) alternating spin chain.

\section{Overlap of $1/6$ BPS Wilson loop}
\label{sec:onesixWL}
In this section, we derive the exact overlap formula for the $1/6$-BPS Wilson loop $|\mathcal{B}^{B,\text{big}}_{1/6}\rangle$ in \eqref{eq:BB16}. We will derive the formula for $|\mathcal{B}_R\rangle$ and $U_{\text{even}}|\mathcal{B}_R\rangle$ separately and then take the sum.\par

We will use the method developed in \cite{Gombor:2020kgu}. For a two-site state $|\mathcal{B}\rangle$ with the selection rule \eqref{selectionRules}, one expects that the overlap takes the following form
\begin{align}
\label{eq:overlapAnsatz}
\dfrac{\langle\mathcal{B}|\textbf{u,v,w}\rangle}{\sqrt{\langle\textbf{u,v,w}|\textbf{u,v,w}\rangle}}=\prod_{j=1}^{K_{\mathbf{u}}}h^{(1)}(u_j)\prod_{k=1}^{K_{\mathbf{w}}/2}h^{(2)}(w_k)\times \sqrt{\dfrac{\text{det}G_+}{\text{det}G_-}}.
\end{align}
where $\det G_{\pm}$ are the Gaudin-like determinants whose definitions were given in \cite{Yang:2021hrl}. The prefactors $h^{(1)}(u)$ and $h^{(2)}(w)$ can be calculated by a nesting procedure in the sparse limit where $L\to\infty$ and the number of excitations are kept finite
\cite{Gombor:2020kgu,Gombor:2020auk,Gombor:2022aqj}. In this limit, the ratio of determinants $\det G_+/\det G_-\to 1$ and we are left with the contribution from the prefactors. However, this method can not be applied directly in the current situation, for the following two reasons.\par

First, the Bajnok-Gombor nesting procedure starts with evaluating the overlap $\langle \mathcal{B}|0\rangle$, where $|0\rangle$ is the pseudovacuum state. However, it is easy to see this overlap is vanishing for our Wilson loop boundary states. Second, from $R$-charge conservation, the overlap $\langle \mathcal{B}|\textbf{u,v,w}\rangle$ for the Wilson loop boundary states are non-zero only if $K_{\mathbf{u}}=K_{\mathbf{v}}=K_{\mathbf{w}}=L$. Therefore we cannot take the limit $L\to\infty$ while keeping the excitation numbers finite.

\subsection{Rotating boundary state}
To address these two issues, one can rotate the boundary state by a certain angle $\theta$~\cite{Gombor:2022aqj}. The $K$-matrix \eqref{BYBE} still satisfies the BYBE under a $SO(4)$ rotation and hence integrability is preserved. The overlap for the rotated boundary state $|\mathcal{B}_{\theta}\rangle$ is no longer constrained by the selection rule $K_{\mathbf{u}}=K_{\mathbf{v}}=K_{\mathbf{w}}=L$ and we can apply Bajnok-Gombor approach to obtain the prefactor. Assuming the $\theta\rightarrow 0$ limit is smooth, we then obtain the prefactors of the original boundary state by taking $\theta=0$. We will see that this method indeed gives the correct result for the $1/6$-BPS Wilson loop.\par

We first consider the boundary state $\langle\mathcal{B}_R|$ in \eqref{boundaryState}
with the following $SO(4)$ rotation \cite{Gombor:2022aqj}
\begin{align}
\label{eq:SO4rotate}
g(\theta)=\left(\begin{array}{cccc}
\cos \theta & 0 & 0 & -\sin \theta \\
0 & \cos \theta & -\sin \theta & 0 \\
0 & \sin \theta & \cos \theta & 0 \\
\sin \theta & 0 & 0 & \cos \theta
\end{array}\right)
\end{align}
and define
\begin{align}
	R(\theta)=g(\theta) R g(-\theta).
\end{align}
The rotated dual boundary state is given by\footnote{For the computation of the overlap, we consider the dual boundary state.}
\begin{align}
\langle\mathcal{B}_{R(\theta)}|=\left({{R(\theta)}^I}_J\langle I,J|\right)^{\otimes L}
\end{align}
Similarly, we define
\begin{align}
\langle \widehat{\mathcal{B}}_{R(\theta)}|=\langle\mathcal{B}_{R(\theta)}|U_{\text{even}}^{\dagger}\,.
\end{align}
We find that
\begin{align}
{R(\theta)^I}_J\langle I,J |=&\,i\cos (2 \theta)\left(\langle 1\bar{1}|+\langle 2\bar{2}|-\langle 3\bar{3}|-\langle 4\bar{4}|\right)\\\nonumber
\phantom{=}&\,+i\sin(2\theta)\left(\langle1\bar{4}|+\langle 2\bar{3}|+\langle 3\bar{2}|+\langle4\bar{1}| \right)
\end{align}
where the second line which breaks the selection rule $K_{\mathbf{u}}=K_{\mathbf{v}}=K_{\mathbf{w}}=L$. The pseudovacuum state is
\begin{align}
|0\rangle=(|1\bar{4}\rangle)^{\otimes L}.
\end{align}
We thus have
\begin{align}
\langle\mathcal{B}_{R(\theta)}|0\rangle=(i\sin 2 \theta)^L,\qquad
&\langle\widehat{\mathcal{B}}_{R(\theta)}|0\rangle=(i\sin 2 \theta)^L.\\\nonumber
\end{align}
We will perform the Bajnok-Gombor procedure for the state $|\mathcal{B}_{R(\theta)}\rangle$, the state $|\widehat{\mathcal{B}}_{R(\theta)}\rangle$ can be treated similarly. We first define the renormalized state
\begin{align}
\label{eq:renormalizedLevel1}
\langle\mathcal{B}_{R(\theta)}^{(1)}|=\frac{\langle\mathcal{B}_{R(\theta)}|}{(i\sin 2 \theta)^L}\,, \qquad\langle\mathcal{B}_{R(\theta)}^{(1)}|0\rangle=1.
\end{align}

\subsection{Two-particle state}
Now we consider the excited state with two particles with rapidities $u$ and $v$. We denote the type of the particles by $a$ and $b$, where $a,b=1,2$. When a type-$a$ sites on an odd (even) site, the field $Y^1$ ($\bar{Y}_4$) is replaced by $Y^{1+a}$ ($\bar{Y}_{4-a}$). It is also possible for two particles with different labels to occupy the same site, leading to the composite excitations $Y^{4}$ on odd sites and $\bar{Y}_{1}$ on even sites. In what follows, we will denote a state with two particles of type-$a$ and -$b$ at sites $2n-1$ and $2m$ by $|2n-1,2m\rangle\!\rangle_{a,b}$. If the two particles are on the same site $n$, we denote the state by $|n\rangle\!\rangle_{\bullet}$. The asymptotic two-particle Bethe state of the SU(4) alternating spin chain has been constructed in \cite{Gombor:2022aqj} and reads
\begin{align}
\label{eq:2ptState}
\begin{aligned}
|\{u\},\{v\}\rangle_{a, b}= & \sum_{m, n=1}^{L} e^{i p n+i q m} \sum_{c, d=1}^{2}(-1)^{d-1} \chi_{a, b}^{c, d}(u-v)|2 n-1,2 m\rangle\!\rangle_{c, d} \\
& +\sum_{n=1}^{L} e^{i(p+q) n}\left(\zeta_{a, b}^{(1)}(u, v)|2 n-1\rangle\!\rangle_{\bullet}+\zeta_{a, b}^{(2)}(u, v)|2 n\rangle\!\rangle_{\bullet}\right)\,.
\end{aligned}
\end{align}
where
\begin{align}
e^{ip}=\frac{u+i/2}{u-i/2},\qquad e^{iq}=\frac{v+i/2}{v-i/2}\,.
\end{align}
The coefficients are given by
\begin{align}
\chi_{a, b}^{c, d}(u)=\left\{\begin{array}{ll}
\delta_{a}^{c} \delta_{b}^{d}, & 2 n-1<2 m \\
R_{a b}^{c d}(u), & 2 n-1>2 m
\end{array}\right.
\end{align}
with
\begin{align}
R_{a b}^{c d}(u)=\frac{u}{u-i} \delta_{a}^{c} \delta_{b}^{d}-\frac{i}{u-i} \delta_{a}^{d} \delta_{b}^{c}
\end{align}
The coefficients for the double occupation factor read
\begin{align}
\zeta_{a b}^{(1)}(u, v)=\epsilon_{a b} \frac{-v+i / 2}{u-v-i}, \qquad \zeta_{a b}^{(2)}(u, v)=\epsilon_{a b} \frac{u+i / 2}{u-v-i} .
\end{align}
Two comments are in order for the two particle state \eqref{eq:2ptState}. Firstly, this state is an eigenstate of the Hamiltonian in the asymptotic sense. This means it is only an eigenstate for $L\to\infty$. Secondly, when constructing the asymptotic two-particle state, in principle we should also take into account the possibilities where the two particles are on two different even or odd sites. Nevertheless such states will not contribute to the overlap and can be ignored.
\subsection{First level nesting}
The first level $K$-matrix is given by
\begin{align}
\label{nestedK}
K_{ab}^{(1)}(u)=\lim _{L \rightarrow \infty} \frac{1}{A^L}\frac{\left\langle\mathcal{B}_{R(\theta)}|\{u\},\{-u\}\right\rangle_{a, b}}{\sqrt{\langle \{u\},\{-u\}|\{u\},\{-u\}\rangle_{a,b}}},
\end{align}
where $A^L=\langle \mathcal{B}_{R(\theta)}|0\rangle=(-i\sin2\theta)^L$. The overlap is given by
\begin{align}
\label{eq:manyA}
\left\langle\mathcal{B}_{R(\theta)}|u,-u\right\rangle_{a, b}=\left(
\begin{array}{cc}
A_{11} & A_{12} \\
A_{21} & A_{22} \\
\end{array}
\right)
\end{align}
Taking $v=-u$, the two-particle state is simplified further to
\begin{align}
\label{eq:twoparticlePair}
|\{u\},\{-u\}\rangle_{a,b}=&\,\sum_{n,m}e^{ip(n-m)}\sum_{c,d=1}^2(-1)^{d-1}\chi_{a,b}^{c,d}(2u)|2n-1,2m\rangle\!\rangle_{c,d}\\\nonumber
&\,+\frac{\epsilon_{ab}}{2}\frac{u+i/2}{u-i/2}\sum_n\left(|2n-1\rangle\!\rangle_{\bullet}+|2n\rangle\!\rangle_{\bullet} \right)
\end{align}
We have
\begin{align}
\label{eq:overlapElements11}
&\langle\mathcal{B}_{R(\theta)}|2m-1,2n\rangle\!\rangle_{1,1}=\langle\mathcal{B}_{R(\theta)}|2m-1,2n\rangle\!\rangle_{2,2}=(i\sin2\theta)^L\,\delta_{m,n}\,,\\\nonumber
&\langle\mathcal{B}_{R(\theta)}|2m-1,2n\rangle\!\rangle_{1,2}=i\cos2\theta(i\sin2\theta)^{L-1}\delta_{m,n}\,,\\\nonumber
&\langle\mathcal{B}_{R(\theta)}|2m-1,2n\rangle\!\rangle_{2,1}=-i\cos2\theta(i\sin2\theta)^{L-1}\delta_{m,n}\,.
\end{align}
and
\begin{align}
\label{eq:doubleOccup}
\langle\mathcal{B}_{R(\theta)}|2n-1\rangle\!\rangle_{\bullet}=-(i\cos2\theta)(i\sin2\theta)^{L-1},\qquad
\langle\mathcal{B}_{R(\theta)}|2n\rangle\!\rangle_{\bullet}=(i\cos2\theta)(i\sin2\theta)^{L-1}\,.
\end{align}
From \eqref{eq:doubleOccup}, it is clear that the boundary state has vanishing overlap with the second line in \eqref{eq:twoparticlePair} for all $a,b=1,2$.
Using \eqref{eq:overlapElements11}, the matrix components of \eqref{eq:manyA} and \eqref{eq:doubleOccup} can be computed straightforwardly, yielding
\begin{align}
&A_{11}=L\,(i\sin2\theta)^L\,,\qquad A_{22}=-L\,(i\sin2\theta)^L\,,\\\nonumber
&A_{12}=A_{21}=-L(i\cos2\theta)(i\sin2\theta)^{L-1}.
 \end{align}
The norm of the Bethe state is more involved, but the $L\to\infty$ the leading term is simply
\begin{align}
\lim_{L \to\infty}\langle\{u\},\{-u\}|\{u\},\{-u\}\rangle_{a,b}=L^2+\cdots
\end{align}
where the ellipsis denote subleading terms. Therefore in the $L\to\infty$ limit, we obtain
\begin{align}
\label{6nestedK}
K_{ab}^{(1)}(u)=\lim_{L\to\infty}\frac{1}{L (-i\sin 2\theta)^L}\left(
\begin{array}{cc}
A_{11} & A_{12} \\
A_{21} & A_{22} \\
\end{array}
\right)=\left(
\begin{array}{cc}
1 & -\cot 2\theta \\
-\cot 2\theta & -1 \\
\end{array}
\right)
\end{align}
for $1/6$-BPS Wilson loop. Following Bajnok-Gombor approach, \eqref{6nestedK} implies that
\begin{align}
h^{(1)}(u)=K^{(1)}_{1,1}(u)=1
\end{align}

\subsection{Second level nesting}
At the second level, we can take a shortcut. It has been shown \cite{Gombor:2020kgu} that for a boundary state described by the following $K$-matrix
\begin{align}
K=\left(\begin{array}{cc} 1 & -i e^{-\gamma}(\cosh \beta+2 \alpha \sinh \beta)\\ -i e^{-\gamma}(\cosh \beta-2 \alpha \sinh \beta) & -e^{-2\gamma}  \end{array}\right)
\end{align}
the absolute value of the prefactor is given by
\begin{align}
\label{eq:genK}
h^{(2)}(w)=e^{-2\gamma} (\sinh\beta)^{2} \frac{w^{2}+\alpha^{2}}{w(w-i/2)}
\end{align}
Comparing \eqref{eq:genK} with \eqref{6nestedK}, we find that they are identical if we take
\begin{align}
\gamma=0,\qquad \alpha=0,\qquad \cosh\beta=-i\cot(2\theta)
\end{align}
Therefore we conclude that
\begin{align}
\label{eq:h2res}
h^{(2)}(w)=-\frac{1}{(\sin2\theta)^2}\frac{w}{w-i/2}\,.
\end{align}
One comment is that in principle at the second level nesting we should consider the inhomogeneous spin chain and the resulting prefactors contains contributions from the inhomogeneities --- Bethe roots from the first level nesting. At the same time, we should renormalize the second level boundary state similar to \eqref{eq:renormalizedLevel1}. This cancels precisely the contributions from the inhomogeneities. Therefore our result is valid.
Plugging $h^{(1)}(u)=1$ and \eqref{eq:h2res} into \eqref{eq:overlapAnsatz}, we obtain the general overlap formula\footnote{This overlap formula can be also obtained using the recursion method in~\cite{Gombor:2021hmj}.}
\begin{align}
\label{eq:rotatedFinal}
\frac{|\langle\mathcal{B}_{R(\theta)}|\mathbf{u},-\mathbf{u},\mathbf{w}\rangle|^2}{\langle \mathbf{u},-\mathbf{u},\mathbf{w}|\mathbf{u},-\mathbf{u},\mathbf{w}\rangle}=(\sin2\theta)^{2(L-K_{\mathbf{w}})}\prod_{i=1}^{K_{\mathbf{w}}/2}\frac{4 w_i^2}{4w_i^2+1}\times\frac{\det G^+}{\det G^-}.
\end{align}
We have $L\le K_{\mathbf{u}}=K_{\mathbf{v}}\le K_{\mathbf{w}}$. From \eqref{eq:rotatedFinal}, we find that the limit
$\theta\rightarrow 0$ is none vanishing only if $L=K_{\mathbf{u}}=K_{\mathbf{v}}=K_{\mathbf{w}}$, in which case we obtain
\begin{align}\label{6formula}
\frac{|\langle\mathcal{B}_{R}|\mathbf{u},-\mathbf{u},\mathbf{w}\rangle|^2}{\langle \mathbf{u},-\mathbf{u},\mathbf{w}|\mathbf{u},-\mathbf{u},\mathbf{w}\rangle}= \prod_{i=1}^{K_{\mathbf{w}}/2}\frac{w_i^2}{w_i^2 +1/4}\times\frac{\det G^+}{\det G^-}\,.
\end{align}
We have tested \eqref{6formula} numerically up to $L=K_{\mathbf{u}}=K_{\mathbf{v}}=K_{\mathbf{w}}=4$, which is already quite non-trivial.\footnote{The Bethe roots used in this numerical test are listed in Appendix~\ref{sc:Betheroots}. We exploit the coordinate Bethe ansatz in~\cite{Yang:2021hrl} to construct the Bethe states.} Notice that the above result is derived for $K_{\mathbf{w}}$ being even. Numerical computation shows that the overlap vanishes for odd $K_{\mathbf{w}}$.

\subsection{Exact overlap for the shifted boundary state}
Now we move to compute the exact overlap for $\langle \widehat{\mathcal{B}}_{R}|=\langle\mathcal{B}_R|U_{\text{even}}^{\dagger}$ where $U_{\text{even}}$ is defined in \eqref{Ueven} and it shifts all the even sites to the left by two units.
We make the same assumption \eqref{eq:overlapAnsatz} about the exact overlap for $\langle\widehat{\mathcal{B}}_R|\mathbf{u},\mathbf{v},\mathbf{w}\rangle$. To determine the prefactors, we need to compute the overlap with the two-particle state
\begin{align}
\langle\widehat{\mathcal{B}}_{R}|\{u\},\{-u\}\rangle_{a,b}=\langle\mathcal{B}_R|U_{\text{even}}^{\dagger}|\{u\},\{-u\}\rangle_{a,b}
\end{align}
From the definition of $U_{\text{even}}$, it is clear that\footnote{The following computation is precise in the asymptotic sense, which  is enough for our purpose.}
\begin{align}
U_{\text{even}}^{\dagger}|\{u\},\{-u\}\rangle_{a,b}=&\,\sum_{n,m}e^{ip(n-m)}\sum_{c,d=1}^{2}(-1)^{d-1}\chi_{a,b}^{c,d}(2u)|2n-1,2(m+1)\rangle\!\rangle_{c,d}\\\nonumber
&\,+\frac{\epsilon_{ab}}{2}\frac{u+i/2}{u-i/2}\sum_n\left(|2n-1\rangle\!\rangle_{\bullet}+|2n\rangle\!\rangle_{\bullet} \right)\\\nonumber
=&\,\left(\frac{u+i/2}{u-i/2}\right)\sum_{n,m}e^{ip(n-m)}\sum_{c,d=1}^{2}(-1)^{d-1}\chi_{a,b}^{c,d}(2u)|2n-1,2m\rangle\!\rangle_{c,d}\\\nonumber
&\,+\frac{\epsilon_{ab}}{2}\frac{u+i/2}{u-i/2}\sum_n\left(|2n-1\rangle\!\rangle_{\bullet}+|2n\rangle\!\rangle_{\bullet} \right)\,.
\end{align}
Namely, after the action of $U_{\text{even}}^{\dagger}$, the first line is multiplied by a global factor while the second line is left invariant. As we have shown that the second line does not contribute to the overlap. Therefore, $\langle\widehat{\mathcal{B}}_{R(\theta)}|\{u\},\{-u\}\rangle_{a,b}$ is simply proportional to $\langle{\mathcal{B}}_{R(\theta)}|\{u\},\{-u\}\rangle_{a,b}$. The corresponding first level $K$-matrix is given by
\begin{align}
\widehat{K}_{ab}^{(1)}(u)=\lim _{L \rightarrow \infty} \frac{1}{A^L}\frac{\left\langle\widehat{\mathcal{B}}_{R(\theta)}|\{u\},\{-u\}\right\rangle_{a, b}}{\sqrt{\langle \{u\},\{-u\}|\{u\},\{-u\}\rangle_{a,b}}}=\left(\frac{u+i/2}{u-i/2}\right)K^{(1)}_{ab}(u)\,.
\end{align}
%
Therefore we arrived at the following exact overlap formula for $|\widehat{\mathcal{B}}_R\rangle$
\begin{align}
\label{eq:relativePhase16}
\frac{\langle\widehat{\mathcal{B}}_{R}|\mathbf{u},\mathbf{v},\mathbf{w}\rangle}{\sqrt{\langle \mathbf{u},\mathbf{v},\mathbf{w}|\mathbf{u},\mathbf{v},\mathbf{w}\rangle}}= \prod_{j=1}^{K_{\mathbf{u}}}\frac{u_j+i/2}{u_j-i/2}\frac{{\langle\mathcal{B}}_{R}|\mathbf{u},\mathbf{v},\mathbf{w}\rangle}{\sqrt{\langle \mathbf{u},\mathbf{v},\mathbf{w}|\mathbf{u},\mathbf{v},\mathbf{w}\rangle}}.
\end{align}
Hence there is a relative phase between these two boundary state.
From this we get
\begin{align}
\frac{|\langle\mathcal{B}_{1/6}^{B, \mathrm{big}}|\mathbf{u},-\mathbf{u},\mathbf{w}\rangle|^2}{\langle \mathbf{u},-\mathbf{u},\mathbf{w}|\mathbf{u},-\mathbf{u},\mathbf{w}\rangle}=\Bigg| 1+\prod_{j=1}^{K_{\mathbf{u}}}\frac{u_j+i/2}{u_j-i/2}\Bigg|^2
\frac{|\langle\mathcal{B}_{R}|\mathbf{u},-\mathbf{u},\mathbf{w}\rangle|^2}{\langle \mathbf{u},-\mathbf{u},\mathbf{w}|\mathbf{u},-\mathbf{u},\mathbf{w}\rangle}\,.
\end{align}

\section{Overlap of 1/2-BPS Wilson loop}
\label{sec:onehalfWL}
We derive the overlap formula for the 1/2-BPS Wilson loop in this section. The procedure is basically the same as for the 1/6-BPS case. There are two types of 1/2-BPS Wilson loops \eqref{eq:halfBPSp} and \eqref{eq:halfBPSm}, we will consider them in the following two subsections.

\subsection{1/2-BPS Wilson loop $|\mathcal{B}_{1/2}^{1+}\rangle$}
Similar to the 1/6-BPS approach, we need to perform an SO(4) rotation in order to apply the Bajnok-Gombor approach. We take the same rotation as in \eqref{eq:SO4rotate} and define
\begin{align}
U(\theta)=g(\theta)Ug(-\theta),\qquad \langle\mathcal{B}_{U(\theta)}|=\left({U(\theta)^I}_J\langle I,J|\right)^{\otimes L},\qquad
\langle\widehat{\mathcal{B}}_{U(\theta)}|=\langle\mathcal{B}_{U(\theta)}|U^{\dagger}_{\text{even}}\,.
\end{align}
More explicitly, we have
\begin{align}
{U(\theta)^I}_J\langle I,J|=i\cos(2\theta)\left(\langle1\bar{1}|-\langle4\bar{4}|\right)-i\left(\langle2\bar{2}|+\langle3\bar{3}| \right)+i\sin(2\theta)\left(\langle1\bar{4}|+\langle4\bar{1}|\right)
\end{align}
We start with the first level nesting. We have
\begin{align}
\langle\mathcal{B}_{U(\theta)}|0\rangle=(i\sin2\theta)^L\,.
\end{align}
We then compute the overlap $\langle\mathcal{B}_{U(\theta)}|\{u\},\{-u\}\rangle_{a,b}$ where $|\{u\},\{-u\}\rangle_{a,b}$ is defined in \eqref{eq:twoparticlePair}. Using
\begin{align}
&\langle\mathcal{B}_{U(\theta)}|2m-1,2n\rangle\!\rangle_{1,1}=\langle\mathcal{B}_{U(\theta)}|2m-1,2n\rangle\!\rangle_{2,2}=0\,,\\\nonumber
&\langle\mathcal{B}_{U(\theta)}|2m-1,2n\rangle\!\rangle_{1,2}=-i(i\sin2\theta)^{L-1}\,\delta_{m,n}\,,\\\nonumber
&\langle\mathcal{B}_{U(\theta)}|2m-1,2n\rangle\!\rangle_{2,1}=i(-i\sin2\theta)^{L-1}\,\delta_{m,n}\,,\\\nonumber
&\langle\mathcal{B}_{U(\theta)}|2n-1\rangle\!\rangle_{\bullet}=-(i\cos2\theta)(i\sin2\theta)^{L-1}\,,\\\nonumber
&\langle\mathcal{B}_{U(\theta)}|2n\rangle\!\rangle_{\bullet}=(i\cos2\theta)(i\sin2\theta)^{L-1}\,,
\end{align}
We find that
\begin{align}
K_{ab}^{(1)}(u)=\lim_{L\to\infty}\frac{1}{L(-i\sin2\theta)^L}\langle\mathcal{B}_{U(\theta)}|\{u\},\{-u\}\rangle_{a,b}
=\frac{\epsilon_{ab}}{\sin2\theta}\,,
\end{align}
which is nothing but the dimer state. For the second level nesting, we can directly apply the result of the dimer state \cite{Gombor:2020kgu}, which leads to
\begin{align}
\frac{|\langle\mathcal{B}_{U(\theta)}|\mathbf{u},-\mathbf{u},\mathbf{w}\rangle|^2}{\langle\mathbf{u},-\mathbf{u},\mathbf{w}|\mathbf{u},-\mathbf{u},\mathbf{w}\rangle}=\frac{(-1)^L}{(\sin2\theta)^{2(K_{\mathbf{w}}-L)}}
\prod_{i=1}^{K_{\mathbf{u}}}\big(u_i^2+\frac{1}{4}\big)\prod_{j=1}^{[K_{\mathbf{w}}/2]}\frac{1}{w_i^2(w_i^2+1/4)}\,\frac{\det G_+}{\det G_-}\,.
\end{align}
Again, we find that in the $\theta\to 0$ limit, we must have $K_{\mathbf{u}}=K_{\mathbf{v}}=K_{\mathbf{w}}=L$ and the finite result reads
\begin{align}
\frac{|\langle\mathcal{B}_{U}|\mathbf{u},-\mathbf{u},\mathbf{w}\rangle|^2}{\langle\mathbf{u},-\mathbf{u},\mathbf{w}|\mathbf{u},-\mathbf{u},\mathbf{w}\rangle}=(-1)^L
\prod_{i=1}^{K_{\mathbf{u}}}\big(u_i^2+\frac{1}{4}\big)\prod_{j=1}^{[K_{\mathbf{w}}/2]}\frac{1}{w_i^2(w_i^2+1/4)}\,\frac{\det G_+}{\det G_-}\,.
\end{align}
This result has been tested numerically. Our numerical results also reveal that this formula is also correct when $L$ is odd although the above derivation was performed for even $L$. This aspect is different from the bosonic $1/6$-BPS case in the previous section.

\paragraph{Shifted state}
We find that the shifted state overlap is again proportional to the un-shifted one as
\begin{align}
\frac{\langle\widehat{\mathcal{B}}_{U}|\mathbf{u},-\mathbf{u},\mathbf{w}\rangle}{\sqrt{\langle \mathbf{u},-\mathbf{u},\mathbf{w}|\mathbf{u},-\mathbf{u},\mathbf{w}\rangle}}= \prod_{j=1}^{K_{\mathbf{u}}}\left(\frac{u_j+i/2}{u_j-i/2}\right)^2\frac{{\langle\mathcal{B}}_{U}|\mathbf{u},-\mathbf{u},\mathbf{w}\rangle}{\sqrt{\langle \mathbf{u},-\mathbf{u},\mathbf{w}|\mathbf{u},-\mathbf{u},\mathbf{w}\rangle}}.
\end{align}
Notice that the phase factor is different from the 1/6-BPS case. We have tested this result numerically up to $L=K_{\mathbf{u}}=K_{\mathbf{v}}=K_{\mathbf{w}}=4$. Naively, one might expect that according to the same argument of 1/6-BPS case, one should obtain the same phase factor \eqref{eq:relativePhase16}. However, generalizing this argument to the Dimer state seems a bit subtle, as $\widehat{K}_{11}=0$ in this case.

Then we have
\begin{align}
\frac{|\langle\mathcal{B}_{1/2}^{1+}|\mathbf{u},-\mathbf{u},\mathbf{w}\rangle|^2}{\langle\mathbf{u},-\mathbf{u},\mathbf{w}|\mathbf{u},-\mathbf{u},\mathbf{w}\rangle}=\Bigg|1+\prod_{j=1}^{K_{\mathbf{u}}}\left(\frac{u_j+i/2}{u_j-i/2}\right)^2\Bigg|^2\frac{|\langle\mathcal{B}_{U}|\mathbf{u},-\mathbf{u},\mathbf{w}\rangle|^2}{\langle\mathbf{u},-\mathbf{u},\mathbf{w}|\mathbf{u},-\mathbf{u},\mathbf{w}\rangle}\,.
\end{align}

\subsection{1/2-BPS Wilson loop $|\mathcal{B}_{1/2}^{4-}\rangle$}
We take the same rotation as in \eqref{eq:SO4rotate} and define
\begin{align}
\tilde{U}(\theta)=g(\theta)\tilde{U}g(-\theta),\qquad \langle\mathcal{B}_{\tilde{U}(\theta)}|=\left({\tilde{U}(\theta)^I}_J\langle I,J|\right)^{\otimes L},\qquad
\langle\widehat{\mathcal{B}}_{\tilde{U}(\theta)}|=\langle\mathcal{B}_{\tilde{U}(\theta)}|\tilde{U}^{\dagger}_{\text{even}}\,.
\end{align}
More explicitly, we have
\begin{align}
{\tilde{U}(\theta)^I}_J\langle I,J|=i\cos(2\theta)\left(\langle 1\bar{1}|-\langle 4\bar{4}|\right)+i\left(\langle 2\bar{2}|+\langle 3\bar{3}|\right)+i\sin(2\theta)\left(\langle 1\bar{4}|+\langle 4\bar{1}|\right)
\end{align}
The rest of the computations are almost identical to the state $\langle\mathcal{B}_{1/2}^{1+}|$. The only difference is that at the first level nesting we have an additional minus sign, this leads to the following relative phase between the two overlaps
\begin{align}
\langle\mathcal{B}_{\tilde{U}}|\mathbf{u},-\mathbf{u},\mathbf{w}\rangle=(-1)^L\langle\mathcal{B}_{{U}}|\mathbf{u},-\mathbf{u},\mathbf{w}\rangle
\end{align}
The norm of the overlap is thus the same as $\langle\mathcal{B}_{1/2}^{1+}|$
\begin{align}
\frac{|\langle\mathcal{B}_{\tilde{U}}|\mathbf{u},-\mathbf{u},\mathbf{w}\rangle|^2}{\langle\mathbf{u},-\mathbf{u},\mathbf{w}|\mathbf{u},-\mathbf{u},\mathbf{w}\rangle}=
(-1)^L\prod_{i=1}^{K_{\mathbf{u}}}\big(u_i^2+\frac{1}{4}\big)\prod_{j=1}^{[K_{\mathbf{w}}/2]}\frac{1}{w_i^2(w_i^2+1/4)}\,\frac{\det G_+}{\det G_-}\,.
\end{align}
\paragraph{Shifted state} The overlap for the shifted state leads to the same phase factor as for $\langle\mathcal{B}_{1/2}^{1+}|$ case
\begin{align}
\frac{\langle\widehat{\mathcal{B}}_{\tilde{U}}|\mathbf{u},-\mathbf{u},\mathbf{w}\rangle}{\sqrt{\langle \mathbf{u},-\mathbf{u},\mathbf{w}|\mathbf{u},-\mathbf{u},\mathbf{w}\rangle}}= \prod_{j=1}^{K_{\mathbf{u}}}\left(\frac{u_j+i/2}{u_j-i/2}\right)^2\frac{{\langle\mathcal{B}}_{\tilde{U}}|\mathbf{u},-\mathbf{u},\mathbf{w}\rangle}{\sqrt{\langle \mathbf{u},-\mathbf{u},\mathbf{w}|\mathbf{u},-\mathbf{u},\mathbf{w}\rangle}}.
\end{align}
So as in the previous case, we have,
\begin{align}
\frac{|\langle\mathcal{B}_{1/2}^{4-}|\mathbf{u},-\mathbf{u},\mathbf{w}\rangle|^2}{\langle\mathbf{u},-\mathbf{u},\mathbf{w}|\mathbf{u},-\mathbf{u},\mathbf{w}\rangle}=\Bigg|1+\prod_{j=1}^{K_{\mathbf{u}}}\left(\frac{u_j+i/2}{u_j-i/2}\right)^2\Bigg|^2\frac{|\langle\mathcal{B}_{\tilde{U}}|\mathbf{u},-\mathbf{u},\mathbf{w}\rangle|^2}{\langle\mathbf{u},-\mathbf{u},\mathbf{w}|\mathbf{u},-\mathbf{u},\mathbf{w}\rangle}\,.
\end{align}
Finally for $|\mathcal{B}_{1/3}\rangle$, we have
\begin{align}
\frac{|\langle\mathcal{B}_{1/3}|\mathbf{u},-\mathbf{u},\mathbf{w}\rangle|^2}{\langle\mathbf{u},-\mathbf{u},\mathbf{w}|\mathbf{u},-\mathbf{u},\mathbf{w}\rangle}&=\Bigg|(n_1+(-1)^Ln_4)\left(1+\prod_{j=1}^{K_{\mathbf{u}}}\left(\frac{u_j+i/2}{u_j-i/2}\right)^2\right)\Bigg|^2\nonumber\\
& \times \frac{|\langle\mathcal{B}_{U}|\mathbf{u},-\mathbf{u},\mathbf{w}\rangle|^2}{\langle\mathbf{u},-\mathbf{u},\mathbf{w}|\mathbf{u},-\mathbf{u},\mathbf{w}\rangle}\,.
\end{align}

\section{Conclusion}
\label{sec:conclude}
In this paper, we computed the correlation function of a circular BPS Wilson loop and a single-trace operator in ABJM theory. This correlation function is proportional to the overlap of a boundary state from the Wilson loop and the Bethe state corresponding to the single-trace operator. We proved that among a sub-class of the fermionic $1/6$-BPS Wilson loops, only two special cases, bosonic $1/6$-BPS Wilson loops and half-BPS Wilson loops can lead to  tree-level integrable boundary states. The boundary state from the $1/3$-BPS Wilson loop is integrable  at tree level as well.  Our result for the subclass of  fermionic $1/6$-BPS Wilson loops is in some sense similar to the results on integrability of the open chains from bosonic (non-)supersymmetric Maldacena-Wilson lines in the $\mathcal{N}=4$ SYM~\cite{Correa:2018fgz}. There it was found that only the two special cases, half-BPS Maldacena-Wilson loops and the usual Wilson loops, lead to integrable open chains.
We  also obtained the exact overlap formulae up to a phase for all the  tree-level integrable boundary states corresponding to $W^B_{1/6}$, $\hat{W}^B_{1/6}$, $W^{B, \mathrm{big}}_{1/6}$, $W_{1/2}^{1+}$, $W_{1/2}^{4-}$ and $W_{1/3}$. 

There are various directions that deserve further investigations.  One immediate problem is to fix the phase of the overlaps for  $W^B_{1/6}$, and $W_{1/2}^{1+}$. Although as long as the Wilson loop one-point function is concerned, the phase is unimportant, as a spin chain problem it is still nice to have a method which also gives the phase factor.
It is valuable to generalize our result to other closed sectors  even to the full sector at two loop level. If some BPS Wilson loops give integrable boundary  states in the full sector, the constraints from the bosonic and fermionic duality~\cite{Kristjansen:2020vbe, Kristjansen:2021xno, Kristjansen:2021abc} of the Bethe ansatz equations may help us to pin down the full sector overlap formulas.
The next step is to  obtain all loop overlap in the asymptotic sense using integrable bootstrap method, as done for some integrable boundary states in $\mathcal{N}=4$ SYM~\cite{Jiang:2019xdz, Jiang:2019zig, Komatsu:2020sup} case. An even more ambitious goal would be computing the finite-size corrections
using the worldsheet $g$-function approach~\cite{Jiang:2019xdz, Jiang:2019zig}.

In this paper, we only consider certain BPS Wilson loop in the fundamental representation of a suitable group or super-group. It is interesting to study whether similar Wilson loops in higher dimensional representations can also lead to
integrable boundary states. Generating functions of Wilson loops in various representations~\cite{Hartnoll:2006is} and {the method of introducing one-dimensional scalars and/or fermions along the contour of the Wilson loop}~\cite{JKV, Gomis:2006sb, Gomis:2006im, Castiglioni:2022yes} should be very helpful here.

One common feature of the Wilson loops considered here is that the scalar coupling is constant.  There are also other
BPS Wilson loops whose scalar couplings are $\tau$-dependent~\cite{Cardinali:2012ru, Bianchi:2014laa, Castiglioni:2022yes}, where $\tau$ is used to parameterise the Wilson loop contour.
The correlator of such Wilson loops and a single-trace  operator in the $SU(4)$ sector will lead to boundary states involving integration of $\tau$ along the circle. It is interesting to seek integrable boundary states among them.

Comparing with the $\mathcal{N}=4$ SYM case, the study of correlators of BPS Wilson loops and single-trace operators is much more preliminary and there are far less results. One direction  complementary to the study here is using localization~\cite{Kapustin:2009kz} to compute the correlation functions of BPS Wilson loops and certain BPS local operators and comparing the results at strong coupling in the large $N$ limit with holographic computations. Some computations in $\mathcal{N}=4$ SYM case in this direction can be found in~\cite{Giombi:2009ds, Giombi:2012ep, Giombi:2020amn, Berenstein:1998ij, Giombi:2006de, Zarembo:2002ph}, however the localization computation in the ABJM case seems more challenging and calls for new developments. Some important progress in this direction was made recently in~\cite{Guerrini:2023rdw}.

\section*{Acknowledgement}
We would like to thank Bin~Chen for collaboration at  early stages of this project, Hong-Fei~Shu, Jiaju~Zhang for help discussions, and Zhi-Xin~Hu for helps on using the computer cluster at CJQS, TJU. Y.J. would like to thank Center for Joint Quantum Studies of Tianjin University for hospitality during the final stage of the work.
The work of Y.J.  is partly supported  by Startup Funding no. 3207022217A1 of Southeast University.
The work of J.W. and P.Y.  is partly supported  by the National Natural Science Foundation of China, Grant No.~11975164, 11935009, 12247103, 12047502,
and  Natural Science Foundation of Tianjin under Grant No.~20JCYBJC00910.

\appendix

\section{The Lagrangian  and supersymmetry transformations of ABJM theory}
\label{sc:lagrangian}

{\bf Spinor convention.} The circular BPS Wilson loops in ABJM theory can only exist when the theory is put in the Euclidean space $\mathbf{R}^3$
~\cite{Ouyang:2015ada}.
We follow the spinor convention in~\cite{Ouyang:2015ada}. The metric on $\mathbf{R}^3$ is $\delta_{\mu\nu}=\mathrm{diag}(1, 1, 1)$, the coordinates are
$x^\mu=(x^1, x^2, x^3)$. The $\gamma$ matrices are
\be  \gamma^{\mu\,\,\,\beta}_{\,\,\alpha}=(-\sigma^2, \sigma^1, \sigma^3)\,.\ee
They satisfy
\be \gamma^\mu\gamma^\nu=\delta^{\mu\nu}+i \epsilon^{\mu\nu\rho} \gamma_\rho\,,\ee
where $\epsilon^{\mu\nu\rho}$  is the rank-$3$  antisymmetric tensor with $\epsilon^{123}=1$. We raise or lower the spinor indices using anti-symmetric tensor $\epsilon^{\alpha\beta}$ and $\epsilon_{\alpha\beta}$
\be \theta^\alpha=\epsilon^{\alpha\beta}\theta_\beta\,, \, \theta_\alpha=\epsilon_{\alpha\beta}\theta^\beta\,,\ee
with $\epsilon^{12}=-\epsilon_{12}=1$.
We will use the shorthand notation,
\be \theta \psi=\theta^\alpha \psi_\alpha\,, \, \theta\gamma^\mu \psi=\theta^\alpha \gamma^{\mu\,\,\, \beta}_{\,\,\alpha} \psi^\beta. \ee

{\bf Field content.}  ABJM theory is the three-dimensional $\mathcal{N}=6$ super-Chern-Simons theory with gauge group $U(N)\times U(N)$. The Chern-Simons levels
are $k$ and $-k$, respectively. Besides the gauge fields $A_\mu, \hat{A}_\mu$ in the adjoint representation of each $U(N)$. The matter fields include four complex  scalars $Y^I$
and four Dirac spinors $\psi_I$ in the bi-fundamental representation of the gauge group. $Y^I$'s are in the $\bf{4}$ representation of R-symmetry group $SU_R(4)$ and $\psi_I$'s
 are in the $\bar{\bf{4}}$ representation.

\paragraph{Lagrangian.}
The Lagrangian of ABJM theory
can be written as the sum of four parts,
\be\mathcal{L}_{\mathrm{ABJM}}=\mathcal{L}_{\mathrm{CS}}+\mathcal{L}_{k} +\mathcal{L}_{k}+\mathcal{L}_Y\,,\ee
with
\bea \mathcal{L}_{\mathrm{CS}}&=&-\frac{k}{4\pi}\epsilon^{\mu\nu\rho} \mathrm{Tr} \left( A_\mu \partial_\nu A_\rho+\frac{2i}{3}A_\mu A_\nu A_\rho-\hat{A}_\mu \partial_\nu \hat{A}_\rho+\frac{2i}{3}\hat{A}_\mu \hat{A}_\nu \hat{A}_\rho \right)\,,\nonumber\\
\mathcal{L}_p&=& \mathrm{Tr} \left(-D_\mu Y^{\dagger}_I D^\mu Y^I+i \psi^{\dagger I} \gamma^\mu D_\mu \psi_I\right)\,,\nonumber\\
\mathcal{L}_p&=&\frac{4\pi^2}{3k^2}\mathrm{Tr}\left(Y^IY^\dagger_IY^JY^\dagger_J Y^K Y^\dagger_K+Y^IY^\dagger_JY^JY^\dagger_K Y^K Y^\dagger_I+4 Y^IY^\dagger_JY^KY^\dagger_I Y^J Y^\dagger_K\right.\nonumber\\
&&\left.-6 Y^IY^\dagger_JY^JY^\dagger_I Y^K Y^\dagger_K\right)\,,\nonumber \\
\mathcal{L}_Y&=&-\frac{2\pi i }{k} \mathrm{Tr}\left( Y^I Y^\dagger_I \psi_J \psi^{\dagger J} -2 Y^I Y^\dagger_J \psi_I \psi^{\dagger J}-Y^\dagger_I Y^I \psi^{\dagger J}\psi_J+2 Y^\dagger_I Y^J \psi^{\dagger I}\psi_J\right.\nonumber\\
&&\left. + \epsilon_{IJKL} Y^I \psi^{\dagger J}Y^K \psi^{\dagger L} -\epsilon^{IJKL} Y^\dagger_I \psi_J Y^\dagger_K \psi_L\right)\,.
\eea
Here the covariant derivatives are defined as
\bea D_\mu Y^I&=&\partial_\mu Y^I+i A_\mu Y^I-i Y^I\hat{A}_\mu\,, \nonumber\\
     D_\mu Y^{\dagger}_I&=&\partial_\mu Y^\dagger_I+i \hat{A}_\mu Y^\dagger_I-i Y^\dagger_I A_\mu\,, \nonumber\\
     D_\mu \psi_I&=&\partial_\mu \psi_I+i A_\mu \psi_I-i \psi_I\hat{A}_\mu\,.\eea
and $\epsilon_{IJKL}, \epsilon^{IJKL}$ is totally anti-symmetric tensor with $\epsilon_{1234}=\epsilon^{1234}=1$.

\paragraph{Supersymmetry transformations.}
The ABJM action is invariant under the following  supersymmetry transformations~\cite{Gaiotto:2008cg, Hosomichi:2008jb, Terashima:2008sy, Bandres:2008ry}:
\bea \delta A_\mu&=&\frac{2\pi}{k}(Y^I\psi^{\dagger J}\gamma_\mu \varepsilon_{IJ}+\bar{\varepsilon}^{IJ}\gamma_\mu\psi_JY^\dagger_I)\,, \\
\delta \hat{A}_\mu&=&\frac{2\pi}{k} (\psi^{\dagger J}Y^I\gamma_\mu\varepsilon_{IJ}+\bar{\varepsilon^{IJ}}Y^{\dagger}_I\gamma_\mu\psi_J)\,,\\
\delta Y^I&=&i\bar{\varepsilon}^{IJ}\psi_J\,, \,\, \delta Y^\dagger_I=i\psi^{\dagger J}\varepsilon_{IJ}\,,\\
\delta \psi_I&=&\gamma^\mu \varepsilon_{IJ}D_\mu Y^J+\vartheta_{IJ}Y^J-\frac{2\pi}{k}\varepsilon_{IJ}(Y^J Y^\dagger_K Y^K-Y^KY^\dagger_K Y^J)\nonumber\\
&&-\frac{4\pi}{k}\varepsilon_{KL}Y^KY^\dagger_I Y^L\,,\\
\delta \psi^{\dagger I}&=&-\bar{\varepsilon}^{IJ}\gamma^\mu D_\mu Y^\dagger_J+\bar{\vartheta}^{IJ}Y^\dagger_J+\frac{2\pi}{k}\bar{\varepsilon}^{IJ}(Y^\dagger_J Y^K Y^\dagger_K-Y^\dagger_K Y^K Y^\dagger_J)\nonumber\\
&&+\frac{4\pi}{k} \bar{\varepsilon}^{KL}Y^\dagger_K Y^I Y^\dagger_L.\eea

The supersymmetry parameters are $\varepsilon_{IJ}=\theta_{IJ}+x^\mu \gamma_\mu \vartheta_{IJ}$ and $\bar{\varepsilon}^{IJ}=\bar{\theta}^{IJ}-\bar{\vartheta}^{IJ}x^\mu \gamma_\mu$. Here $\theta$'s give the Ponicar\'e supersymmetry, and $\vartheta$'s give the conformal supersymmetry. They satisfy the following constraints,
\bea \theta_{IJ}&=&-\theta_{JI}\,, \, \bar{\theta}^{IJ}=\frac12\epsilon^{IJKL}\theta_{KL}\,,\\
\vartheta_{IJ}&=&-\vartheta_{JI}\,, \bar{\vartheta}^{IJ}=\frac12\epsilon^{IJKL}\vartheta_{KL}\,.\eea
Notice that for the theory in the Euclidean signature we do not impose that $\bar{\theta}^{IJ}$ ($\bar{\vartheta}^{IJ}$) is the complex conjugation of $\theta_{IJ}$ ($\vartheta_{IJ}$) \cite{Nicolai:1978vc}.

\paragraph{Propagators of the scalar fields.}
The tree-level propagators of the scalar fields are,
\be \langle Y^{I\alpha}_{\,\,\,\,\,\,\,\,\bar{\beta}}(x) Y^{\dagger\,\,\bar{\gamma}}_{J\,\,\,\,\,\,\rho}(y)\rangle =\frac{\delta^I_J\delta^\alpha_\rho \delta^{\bar{\gamma}}_{\bar{\beta}}}{4\pi|x-y|}\,,\label{propagator}\ee
where $\alpha, \bar{\beta}, \bar{\gamma}, \rho$ are color indices.

\section{Numerical solutions of the Bethe  equations}
\label{sc:Betheroots}
In this appendix, we present a collection of numerical solutions for the Bethe equations in the $SU(4)$ sector of the ABJM theory
\bea
1=e^{i \phi_{u_{j}}}=\left(\frac{u_{j}+\frac{i}{2}}{u_{j}-\frac{i}{2}}\right)^{L} \prod_{\substack{k=1 \\
		k \neq j}}^{K_{\mathrm{u}}} S\left(u_{j}, u_{k}\right) \prod_{k=1}^{K_{\mathrm{w}}} \tilde{S}\left(u_{j}, w_{k}\right)\,, \\
1=e^{i \phi_{w_{j}}}=\prod_{\substack{k=1 \\
		k \neq j}}^{K_{\mathrm{w}}} S\left(w_{j}, w_{k}\right) \prod_{k=1}^{K_{\mathrm{u}}} \tilde{S}\left(w_{j}, u_{k}\right) \prod_{k=1}^{K_{\mathrm{v}}} \tilde{S}\left(w_{j}, v_{k}\right)\,, \\
1=e^{i \phi_{v_{j}}}=\left(\frac{v_{j}+\frac{i}{2}}{v_{j}-\frac{i}{2}}\right)^{L} \prod_{\substack{k=1 \\
		k \neq j}}^{K_{\mathrm{v}}} S\left(v_{j}, v_{k}\right) \prod_{k=1}^{K_{\mathrm{w}}} \tilde{S}\left(v_{j}, w_{k}\right)\,,
\eea
where the S-matrices  $S(u, v)$  and  $\tilde{S}(u, v)$ are given by
\be
S(u, v) \equiv \frac{u-v-i}{u-v+i}, \quad \tilde{S}(u, v) \equiv \frac{u-v+\frac{i}{2}}{u-v-\frac{i}{2}}\,.
\ee
Here the numbers of rapidities $\mathbf{u,v,w}$ are denoted by $K_u,K_v,K_w$.

The cyclicity property of the single trace operator is equivalent to the zero momentum condition
\be
1=\prod_{j=1}^{K_{\mathbf u}}\frac{u_{j}+\frac{i}{2}}{u_{j}-\frac{i}{2}}\prod_{j=1}^{K_{\mathbf v}}\frac{v_{j}+\frac{i}{2}}{v_{j}-\frac{i}{2}}\,.
\ee

In the following, we present a collection of solutions that fulfill both Bethe ansatz equations and the zero momentum condition. Rational Q-system  \cite{Marboe:2016yyn, Gu:2022dac} plays an important role here.

\newpage
{\centering
	\begin{longtable}{p{0.3cm}|p{0.6cm}|p{13cm}}
		\hline
		\hline
		
		$L$& $K_{\bf w}$ & $[{\bf u}, {\bf v}, {\bf w}]$ \\
		\hline
		1 &1& $[ \{0\}, \{0\},\{ 0\}]$ \\
		\hline
		2 &2& $[\{\sqrt{\tfrac{3}{20}},-\sqrt{\tfrac{3}{20}}\},\{-\sqrt{\tfrac{3}{20}},\sqrt{\tfrac{3}{20}}\},\{\tfrac{1}{\sqrt{5}},-\tfrac{1}{\sqrt{5}}\}]$\\
		\hline
		3&3&$[\{ -0.61842989257770833, \, 0\,, 0.61842989257770833\},$
	
$ \{ 0.61842989257770833, \, 0\, ,  -0.61842989257770833\},$
	
$\{0.71410132990930250,  -0.71410132990930250, 	 0\}] 	$ \\
\hline
&&$[\{\,0.36628143864284446,\, -0.18314071932142223- 0.5006211833519472 i$\\
&&$-0.18314071932142223 + 0.5006211833519472 i\},$\\
&& $ \{\,-0.36628143864284446,\, 0.18314071932142223+0.5006211833519472 i$\\
&& $0.18314071932142223 -0.5006211833519472 i\},$\\
&&  $\{\,-0.4472135954999579 i\,,\,0\,,\,0.4472135954999579i\}] $ \\
	\hline
	&&$[\{\,-0.9393910431943004 i,\,0\, , \,0.9393910431943004 i\},$

	$ \{\,0.9393910431943004 i,\,0\, , \,-0.9393910431943004 i\},$
	
	$\{\,-1.0847153433251056i\,,\,0\,,\,1.0847153433251056i\}] 	$  \\
	\hline
	4 &4& $[\{\,-0.30330564928014186-0.4984162134634997i,\,-0.30330564928014186+0.4984162134634997i\, , \,0.02628462005210284\, ,0.5803266785081809\},$\\
	&&$\{\,0.30330564928014186+ 0.4984162134634997i,\,0.30330564928014186-0.4984162134634997i\, , \,-0.02628462005210284\, ,-0.5803266785081809\},$\\
	&&$\{\,-0.5162715680301216\,,\,-0.5001222335995396 i\,,\,0.5162715680301216\,$\\
	&&$0.5001222335995396 i \}] $  \\
	\hline
	&& $[\{\,-0.16030976462353738-0.9768494810075854 i,\,-0.16030976462353738+0.9768494810075854i\, , \,-0.14056546652006302\, ,0.46118499576713784\},$\\
&&$\{\,0.16030976462353738+0.9768494810075854 i,\,0.16030976462353738-0.9768494810075854i\, , \,0.14056546652006302\, ,-0.46118499576713784\},$\\
&&$\{\-0.2416681458566768\,,\,-1.0684026594894636 i\,,\,0.24166814585667684\,$\\&&$1.0684026594894636 i \}] 	$ \\
	\hline
	&& $[\{\,-0.7905846950242429,\,-0.18184585032628545\, , $\\&&$0.18184585032628545\, ,0.7905846950242429\},$\\
	
	&&$\{\,0.7905846950242429,\,0.18184585032628545\,, $\\&&$-0.18184585032628545\, ,-0.7905846950242429\},$\\
&&$\{\,-0.9018353804885377\,,\,-0.25327661600652046\,,$\\
&&$0.9018353804885376\,,\,0.25327661600652046 \}] $  \\
	\hline
	&&$[\{\,-0.4913865158293109 - 0.5254890261600584i,\,-0.4913865158293109 + 0.5254890261600584i\, , \,0.49138651582931087 - 0.5254890261600584 i\, $\\&&$,0.49138651582931087+ 0.5254890261600584i\},$\\
&&$\{\,0.4913865158293109+0.5254890261600584i,\,0.4913865158293109-0.5254890261600584i\, , \,-0.49138651582931087+ 0.5254890261600584 i\, $\\&&$,-0.49138651582931087-0.5254890261600584i\},$\\
	&&$\{\-0.53872049128905-0.5800492329412736i\,,\,-0.538720491289058+0.5800492329412736i\,,\,0.538720491289058+ 0.5800492329412736i\,,$\\&&$\,0.538720491289058-0.5800492329412736i \}] 	$ \\
	\hline
	&& $[\{\,-0.14792308614892888i,\,0.14792308614892888i\, $\\&&$, \,-0.8757024703174519i\, ,0.8757024703174519i\},$\\
&&$\{\,0.14792308614892888i,\,-0.14792308614892888i\,$\\&&$ , \,0.8757024703174519i\, ,-0.8757024703174519i\},$\\
	&&$\{\-0.1144398362148366\,,\-1.0318645990119506i\,,$\\&&$\,0.1144398362148366\,,\,1.0318645990119506i \}] 	$  \\
\hline
	\end{longtable}
	\par}
Notice that all of the above sets of Bethe roots satisfy the selection rules $K_{\mathbf{u}}=K_{\mathbf{v}}=K_{\mathbf{w}}=L$ and (\ref{selectionRules}).

Moreover, we have also found a set of  Bethe roots with $L=3, K_{\mathbf{u}}=K_{\mathbf{w}}=1, K_{\mathbf{w}}=2$ that satisfies the zero momentum condition and Bethe equations, however it does not satisfy the selection rules $K_{\mathbf u}=K_{\mathbf v}=K_{\mathbf w}=L$ and ~(\ref{selectionRules}). This set of Bethe roots is
\bea
 &u_1=0.866025\,\qquad w_1= 0.866025,\\ &v_1=-0.198072\,\qquad v_2= 0.631084\,.\nonumber
\eea
\printindex


\providecommand{\href}[2]{#2}\begingroup\raggedright\endgroup

\end{document}